\documentclass[letterpaper,twocolumn,10pt]{article}
\usepackage{usenix}
\usepackage[available]{usenixbadges}
\usepackage{cite}
\usepackage{xspace}
\usepackage{subfig, graphicx}
\usepackage{xurl}
\usepackage{hyperref}
\usepackage{marvosym}

\usepackage{tikz}
\usepackage{amsmath}
\usepackage{listings}
\usepackage[ruled,vlined, linesnumbered]{algorithm2e}
\usepackage{bm}
\usepackage{amssymb}
\usepackage{enumitem}
\usepackage{array}
\usepackage{multirow}
\usepackage{booktabs}
\usepackage{threeparttable}

\newcommand{\para}[1]{%
  \ifhmode\\\fi
  \noindent\textbf{#1}%
}
\newcommand{\sys}{\textsc{vCause}\xspace}
\newcommand{\dtree}{DIM-Tree\xspace}
\newcommand{\dtrees}{DIM-Trees\xspace}
\newcommand{\first}{\textsf{(i)}\xspace}
\newcommand{\second}{\textsf{(ii)}\xspace}
\newcommand{\third}{\textsf{(iii)}\xspace}
\newcommand{\fourth}{\textsf{(iv)}\xspace}
\newcommand{\ie}{\emph{i.e.,}\xspace}
\newcommand{\eg}{\emph{e.g.,}\xspace}

\newtheorem{myTheo}{\textbf{Theorem}}
\newtheorem{myDef}{\textbf{Definition}}
\SetKw{Break}{break}
\SetKw{KwAnd}{and}
\SetKw{KwOr}{or}
\newcounter{phase}
\newcommand{\phreset}{\setcounter{phase}{0}}
\newcommand{\phitem}[1]{%
  \stepcounter{phase}
  \ifhmode\par\fi
  \textit{\thephase)\ #1}%
}

\makeatletter
\DeclareRobustCommand{\corrauthmark}{%
  \textsuperscript{\hyperref[corrauth]{\@fnsymbol{1}}}%
}
\makeatother

\begin{document}

\title{\sys: Efficient and Verifiable Causality Analysis for Cloud-based \\ Endpoint Auditing}


\author{
{\rm Qiyang Song$^{1,2}$, Qihang Zhou$^{1,2,}$\thanks{Corresponding authors.\protect\label{corrauth}}, Xiaoqi Jia$^{1,2,}$\corrauthmark, Zhenyu Song$^{1,2}$}\\
{\rm Wenbo Jiang$^{3}$, Heqing Huang$^{5}$, Yong Liu$^{4}$, Dan Meng$^{1,2}$}\\[0.5ex]
$^{1}$Institute of Information Engineering, Chinese Academy of Sciences\\
$^{2}$School of Cyber Security, University of Chinese Academy of Sciences\\
$^{3}$University of Electronic Science and Technology of China \\
$^{4}$Qi An Xin Technology Group Inc., $^{5}$Independent Researcher \\
\{songqiyang, zhouqihang, jiaxiaoqi, songzhengyu\}@iie.ac.cn\\
wenbo\_jiang@uestc.edu.cn, \{heqing.state,mengdan255\}@gmail.com, liuyong03@qianxin.com\\[0.5ex]
}


\maketitle

\begin{abstract}
In cloud-based endpoint auditing, security administrators often rely on the cloud to perform causality analysis over log-derived versioned provenance graphs to investigate suspicious attack behaviors. However, the cloud may be distrusted or compromised by attackers, potentially manipulating the final causality analysis results. Consequently, administrators may not accurately understand attack behaviors and fail to implement effective countermeasures. This risk underscores the need for a defense scheme to ensure the integrity of causality analysis. While existing tamper-evident logging schemes and trusted execution environments show promise for this task, they are not specifically designed to support causality analysis and thus face inherent security and efficiency limitations.

This paper presents \sys, an efficient and verifiable causality analysis system for cloud-based endpoint auditing. \sys integrates two authenticated data structures: a graph accumulator and a verifiable provenance graph. The data structures enable validation of two critical steps in causality analysis: \first querying a point-of-interest node on a versioned provenance graph, and \second identifying its causally related components. Formal security analysis and experimental evaluation show that \sys can achieve secure and verifiable causality analysis with only <\,1\% computational overhead on endpoints and 3.36\% on the cloud.
\end{abstract}



\section{Introduction}
Cloud-based endpoint auditing~\cite{han2020unicorn,hossain2020combating,dong2023we,hassan2020tactical,SymantecEDR,CrowdstrikeEDR} has become a vital security infrastructure for enterprises. It collects system event logs from enterprise endpoints and leverages cloud computing to perform large-scale analysis, particularly in investigating advanced persistent threats (APTs)~\cite{APTNotes}. A key log analysis technique is causality analysis~\cite{fang2022back,hossain2018dependence,liu2018towards}, typically performed on (versioned) provenance graphs~\cite{li2021threat} derived from system logs. It correlates causality dependencies within these graphs, enabling security administrators to trace the information flow of suspicious entities and assess their impact. Maintaining the integrity of causality analysis is critical, as it empowers administrators to fully understand attack behaviors and deploy effective countermeasures.

However, the integrity of causality analysis cannot be fully ensured in a cloud-based auditing infrastructure. Typically, attackers who have infiltrated the infrastructure may alter stored logs to erase traces of their activities, rendering the associated causality analysis incomplete. This threat is well documented---recent reports show that 72\% of security analysts have encountered log tampering~\cite{72per_tampering,enterprise_log_tampering}. Beyond external attacks, the cloud itself may be untrustworthy~\cite{zhou2021veridb,chai2012verifiable} in preserving the integrity of logs and associated causality analysis. For instance, when managing large volumes of daily endpoint logs (\eg 50 GB from 100 endpoints~\cite{fei2021seal}), the cloud may silently discard a portion of the logs due to misconfiguration or self-interest in conserving resources. In summary, these threats lead to incomplete causality analysis, hindering accurate and comprehensive attack investigation.

In practice, while fully preventing this integrity issue may be infeasible, it is vital to develop a practical defense that allows third parties to verify the integrity of causality analysis in the cloud. To our knowledge, no existing solutions are specifically designed for this purpose. Although tamper-evident logging schemes~\cite{paccagnella2020custos,ma2009new,ahmad2022hardlog,laurie2014certificate} and trusted execution environments (TEEs)~\cite{paju2023sok} show promise in offering partial support, they still face the following challenges.

\textbf{Challenge \#1}. Existing tamper-evident endpoint logging schemes \cite{paccagnella2020custos,ma2009new,ahmad2022hardlog} can secure log collection at endpoints and support validation of all produced logs. However, they do not inherently support causality analysis validation in the cloud. Extending these schemes for such validation typically poses significant efficiency challenges: the administrators should retrieve and verify the \emph{entire set} of endpoint logs from the cloud, reconstruct the provenance graph, and re-run the causality analysis to confirm the results. While deploying TEEs in the cloud can offload this burden from administrators, providing efficient causality analysis still remains challenging, as processing massive logs in enclaves often incurs substantial enclave-to-nonenclave context switch overheads.

\textbf{Challenge \#2}. To improve efficiency, certificate transparency~\cite{laurie2014certificate} can be employed for flexible validation of individual logs. With this approach, administrators can selectively retrieve and verify only the subset of logs relevant to a given causality analysis query, and then examine the corresponding results. However, this approach introduces a \emph{completeness problem}: administrators cannot ascertain whether the retrieved logs---or the resulting causality analysis results---are complete with respect to the original query.

In this paper, we present \sys, an efficient and verifiable causality analysis system for cloud-based endpoint auditing. Unlike existing tamper-evident logging schemes, \sys enables direct validation of causality analysis on versioned provenance graphs. It employs two new authenticated data structures: a \emph{graph accumulator} and a \emph{verifiable versioned provenance graph}, which provide cryptographic proofs for two analysis steps: \first querying a point-of-interest node, \ie a version of a suspicious entity at a specific time; and \second identifying the node’s causally related components.

To support proofs for any node in versioned provenance graphs, our accumulator builds on indexed Merkle trees~\cite{liu2021merkle,smith2020dynamic} for node storage and extends them to meet causality analysis requirements. Notably, a standard Merkle tree supports only single-keyword queries, whereas causality analysis requires two: \emph{a system entity ID} and \emph{a timestamp}. To address this, the accumulator organizes Merkle trees hierarchically. Locally, it employs multiple indexed Merkle trees, each storing version nodes for a single system entity and indexing them by creation timestamps, enabling timestamp-based proofs. Globally, it uses a Merkle tree to aggregate these local trees and indexes them by entity IDs, enabling entity-based proofs. By combining proofs from both levels, the accumulator provides complete proof for any node query.

Our verifiable versioned provenance graph provides cryptographic proofs for each node’s causally related components---namely, the nodes and edges along its incoming and outgoing paths. This is achieved via a recursive hash mechanism that computes two digests for each node: an \emph{incoming path digest} and an \emph{outgoing path digest}, which capture the structure of the node’s paths and jointly serve as proofs of its causally related components. Notably, as the graph should continuously incorporate new nodes to encode real-time system events, the outgoing path structures of many nodes may often change, necessitating frequent updates to their outgoing digests. To mitigate this overhead, we combine graph segmentation strategies with the recursive hash mechanism to construct update-efficient outgoing path digests (\S\,\ref{sec:seg_outgoing_path_digests}).

To enable \emph{comprehensive validation} for causality analysis results, \sys integrates two authenticated data structures in a coordinated manner: the verifiable versioned provenance graph encodes a continuous stream of system event logs, while the graph accumulator stores graph nodes,  maintains the global graph digest, and tracks node updates. Consequently, by retrieving node proofs from the accumulator, we can verify the integrity of the initially queried node in causality analysis. Using the node’s internal incoming and outgoing path digests, we can then validate its causally related components.

We formalize the security of \sys and prove its resilience against an adaptive adversary that can modify any part of the causality analysis results. To demonstrate efficiency, we implement a \sys prototype with 3{,}500 lines of C++ code and evaluate it on large-scale public log datasets~\cite{manzoor2016fast,OpTC}. Our results show that \sys processes 25 million logs in 2 minutes and generates a proof for a 100{,}000-node causality analysis in 49\,ms. We further assess runtime overhead in real-world settings by integrating \sys with realistic endpoint loggers and co-deploying it with three common benchmarks under high workloads. Overall, \sys can incur $<\,1\%$ overhead on endpoints and 3.36\% on the cloud.

\para{Contributions.} We make the following key contributions:
\begin{itemize}
\item We propose a verifiable causality analysis system for cloud-based endpoint auditing, enabling third-party verifiers to efficiently validate causality analysis results.

\item We propose a versioned provenance graph that can encode causality relations in system events while providing proofs for each node's causally related components.

\item We propose a graph accumulator based on hierarchical indexed Merkle trees, enabling proof generation for graph node queries. To support dynamic node updates, we further extend it with a dynamic indexed Merkle structure.

\item We implement a \sys prototype and evaluate its security and performance, showing that \sys provides secure, verifiable causality analysis with low overhead. 

\end{itemize}
\section{Background and Related Work}
\label{sec:background}
\subsection{Causality Analysis on Provenance Graphs}
\label{sec:provenance_background}
\begin{figure}[ht]
    \centering
    \includegraphics[width=3in]{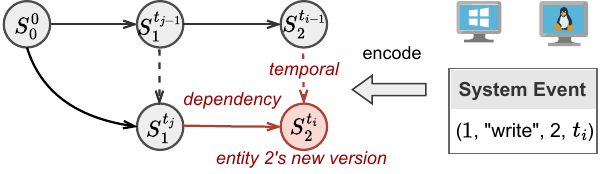}
    \caption{A versioned provenance graph} 

    \label{fig:versioned_provenance_graph}
\end{figure}

\para{Provenance Graphs} record system events and related causal relationships, where nodes represent system entities (\eg processes, files, sockets) and edges represent dependencies among entities (\eg read). They are widely used for causality analysis~\cite{altinisik2023provg,hossain2018dependence} and attack detection~\cite{cheng2024kairos,han2020unicorn,jia2024magic}. However, conventional provenance graphs lack temporal ordering, limiting their ability to accurately reflect causality. To address this, \emph{versioned provenance graphs}~\cite{li2021threat} encode event sequences by creating a new version of an entity whenever its state changes. As shown in Figure~\ref{fig:versioned_provenance_graph}, when event $(1,\text{``write''},2,t_i)$ occurs at timestamp $t_i$, a new node $S_2^{t_i}$ is created to represent the updated version of entity $2$, along with a temporal and dependency edge that capture both temporal and causal relations.

\para{Causality Analysis} is proposed to analyze causality relationships on (versioned) provenance graphs. It was initially introduced by King et al.~\cite{king2003backtracking} to trace the root causes of suspicious system activities, and then extended to file system forensics and intrusion recovery~\cite{goel2005taser,liu2018towards,sitaraman2005forensic}. Building on this foundation, subsequent research has proposed causality analysis variants~\cite{altinisik2023provg,fang2022back,hassan2020tactical,hossain2020combating,hossain2018dependence,ma2016protracer,anjum2022anubis,liu2018towards,hassan2019nodoze,li2023nodlink} to minimize unnecessary logs for manual analysis.

\subsection{Related Integrity Validation Approaches}
\label{sec:related_validation}
\para{Log Integrity Validation}. Prior work has employed trusted execution environments (TEEs)~\cite{sabt2015trusted,karande2017sgx,paccagnella2020custos} (\eg \textsc{CUSTOS}) and cryptographic approaches~\cite{hoang2022faster} to provide tamper evidence for endpoint logs. However, these methods are ill-suited for validating causality analysis results in the cloud. Typically, extending them for such validation requires retrieving full endpoint logs, verifying their integrity, reconstructing the provenance graph, and re-running the analysis, which together incur substantial overhead. Alternatively, other studies have explored specialized hardware~\cite{WORM,ahmad2022hardlog} (\eg \textsc{HardLog}) to protect log integrity, but such solutions are costly and inherently lack support for cloud-side causality analysis.

\para{Graph Integrity Validation}. Prior work has proposed various graph integrity validation schemes~\cite{arshad2017efficient,reina2018method,syalim2010preserving}. While applicable to provenance graphs, they lack support for time-fuzzy node queries essential to causality analysis. Moreover, verifying the neighborhood of a node typically requires either costly upfront commitments (\eg \emph{signing and synchronizing all nodes}) or expensive validation of the entire graph. These schemes also offer limited support for graph updates. 

\para{Other Validation Approaches.} Other related studies focus on verifiable and authenticated provenance storage~\cite{ahmad2020discrepancy,luo2018tprov,jamil2018secure,ruan2019fine}, but do not support causality analysis with verifiable proofs. Alternatively, generic SNARK schemes~\cite{ben2013snarks,bunz2020transparent,chen2022review} could be leveraged to enable such analysis, but large-scale log data make the arithmetic circuits prohibitively large and expensive. Likewise, deploying TEEs~\cite{paju2023sok} in the cloud may incur substantial runtime overhead, as processing massive log volumes within enclaves involves frequent I/O operations and enclave-to-nonenclave context switches.

\section{System Overview}

\subsection{Threat Model}
This work considers a cloud-based endpoint auditing infrastructure in enterprise networks, comprising a set of endpoints, a remote security administrator, and a cloud service. The administrator relies on the cloud to manage system-level event logs from endpoints and to perform causality analysis for attack investigation. According to standard threat models in cloud computing~\cite{zhou2021veridb,wang2023locality}, security risks in such settings often arise from the untrusted cloud. Sophisticated attackers who have infiltrated the infrastructure may tamper with raw logs or directly alter causality analysis results on the cloud, erasing traces of their malicious activities~\cite{enterprise_log_tampering}. While existing log protection schemes~\cite{ahmad2022hardlog,paccagnella2020custos,karande2017sgx,hoang2022faster} can secure log collection, caching, and processing at endpoints, they are not inherently designed to protect logs or validate causality analysis results in the cloud. In this work, we build upon existing log protection schemes and focus on addressing the integrity issues of stored logs and causality analysis in the cloud.

Beyond external threats, the cloud itself may be untrustworthy in maintaining the integrity of logs and associated causality analysis results, due to misconfiguration or self-interest~\cite{chai2012verifiable,zhou2021veridb}. For instance, when handling large volumes of event logs with limited configured resources, it may delete logs to free storage or return incomplete analysis results to conserve computational resources. To account for these risks, we assume the cloud may exhibit Byzantine faults~\cite{zhou2021veridb}, meaning it can: \first manipulate system event logs by inserting, modifying, or deleting arbitrary data; and \second return incomplete, outdated, or falsified causality analysis results. Our work focuses on the integrity issues rather than confidentiality, as the latter is orthogonal to our goals and has been addressed by existing privacy-preserving graph search schemes~\cite{cao2011privacy,wang2022pegraph}.

\subsection{Problem Formulation}
In this work, we build versioned provenance graphs from system logs to support verifiable causality analysis. 
We focus on providing validation for the generic form of causality analysis. Additional  variants~\cite{ma2016protracer,liu2018towards,hassan2019nodoze,fang2022back,li2023nodlink} that prune unnecessary analysis paths are also supported, as their outputs are subsumed by the general causality analysis and thus covered by our validation~(see \S\,\ref{sec:discussion}). Following prior work~\cite{fang2022back,fei2021seal,hossain2018dependence}, the causality analysis can be formulated as a subgraph query over a versioned provenance graph, as follows:

\begin{myDef}[Causality Analysis]
\label{def:causality_analysis}
This process begins with a node query $N(s, \preceq t)$ to locate the point-of-interest version node $n$, whose system entity ID matches $s$ and whose creation timestamp is the nearest one preceding or equal to $t$. Subsequently,         backward and forward depth-first searches are performed to extract the backward and forward causally related components, i.e., $\{V_{\rightarrow n}, E_{\rightarrow n}\}$ and $\{V_{n \rightarrow}, E_{n \rightarrow}\}$:
\begin{equation}
\label{eq:backward_causality}
    V_{\rightarrow n} = \{ i \mid Path_{i, n} \neq \varnothing \},  E_{\rightarrow n} = \{ i.E_{in} \mid Path_{i, n} \neq \varnothing \}
\end{equation}
\begin{equation}
\label{eq:forward_causality} 
    V_{ n \rightarrow} = \{ j \mid Path_{n, j} \neq \varnothing \}, 
    E_{n \rightarrow} = \{ j.E_{out} \mid Path_{n, j} \neq \varnothing \}
\end{equation}
Here, $Path_{i, j}$ denotes paths between node $i$ and $j$, and $E_{in}$ and $E_{out}$ refer to a node's incoming and outgoing edges. 
\end{myDef}
By definition, node $n$'s forward and backward causally related components (\ie $\{V_{ \rightarrow n}, E_{\rightarrow n}\}$ and $\{V_{ n \rightarrow}, E_{n \rightarrow}\}$) are actually the nodes and edges along $n$'s incoming and outgoing paths.

\para{Causality Analysis Validation.} Accordingly, verifying a causality analysis result should  \first check the correctness of the initially queried node, and \second validate the integrity of forward and backward causally related nodes and edges.

\subsection{System Model and Workflow}

\begin{figure}[t]
    \centering
\includegraphics[width=3.2in]{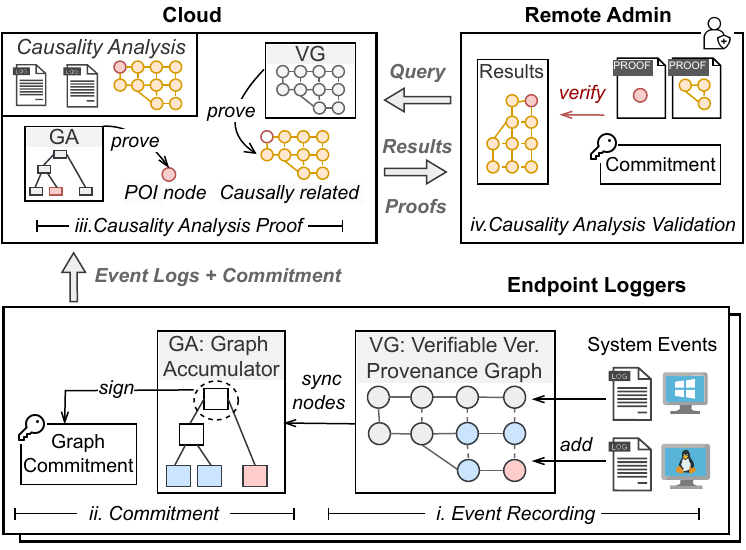}
    \caption{Architecture of \sys} 
    \label{fig:overview}
\end{figure}

Figure \ref{fig:overview} shows the architecture of \sys, an efficient and verifiable causality analysis system for cloud-based endpoint auditing. Overall, \sys interacts with three parties: (i) \emph{endpoint loggers}, which upload real-time system event logs to the cloud; (ii) the \emph{cloud}, which manages endpoint logs and provides a verifiable causality analysis service; (iii)  \emph{a remote security administrator}, who performs causality analysis to investigate suspicious system activities. 

To achieve causality analysis validation, \sys adopts two authenticated data structures: a \emph{graph accumulator} and a \emph{verifiable versioned provenance graph structure}. The structures respectively provide proofs for the initially queried point-of-interest (POI) node and its causally related components.

\para{Graph Accumulator.} 
To provide proofs for node queries on a versioned provenance graph, a naive solution is to use an indexed Merkle tree~\cite{liu2021merkle,smith2020dynamic} for node storage. However, a single tree only supports single-keyword queries, while causality analysis typically requires two: a system entity ID and a timestamp. Our accumulator addresses this with a \emph{hierarchical Merkle tree} design. Locally, each system entity is assigned an indexed Merkle tree that stores its versioned nodes by timestamp, enabling timestamp-based proofs. Globally, a Merkle tree aggregates these local trees by entity ID, enabling entity-based proofs. By combining local and global proofs, the accumulator produces complete node-query proofs. To support graph evolution, we also design a dynamic Merkle tree structure that supports efficient node insertions.

\para{Verifiable Versioned Provenance Graph.}
The graph structure extends the original provenance graph, capturing causality relations in system events while also providing verifiable proofs for a node's causally related components. According to Equation \ref{def:causality_analysis}, a node’s causally related components are located in its incoming and outgoing paths. To enable their validation, we employ a recursive hash mechanism that computes two digests for each node: the incoming and outgoing path digests. These digests capture the structural information of a node’s incoming/outgoing paths and jointly serve as proofs of its causally related components. 
Notably, as the graph continuously incorporates new nodes to encode real-time events, the outgoing path structures of many existing nodes may change frequently, necessitating frequent updates to their outgoing path digests. To mitigate this overhead, we adopt \emph{segmented outgoing path digests}, generated based on segmentation strategies that partition the graph into smaller dependency trees. This ensures that adding a new node affects only the nodes within a single tree, significantly reducing the number of required digest updates.

Based on the structures, \sys operates in four phases:
\phreset
 \phitem{Event Recording:} Each endpoint logger continuously captures system events and updates the corresponding verifiable versioned provenance graph to encode these events.
\phitem{Commitment:} Each endpoint logger periodically synchronizes node changes during event recording into the Merkle tree of the graph accumulator, signs the tree root as a graph commitment, and sends it along with event logs to the cloud.
\phitem{Causality Analysis Proof:} The cloud uses event logs from each endpoint to reconstruct graph accumulators and verifiable versioned provenance graphs. Upon receiving a request from the administrator, it performs causality analysis and generates corresponding proofs using these structures.
\phitem{Causality Analysis Validation:} By retrieving the graph commitment and node proofs from the cloud, the administrator first verifies the integrity of the initially queried node. Then, using the node’s incoming and outgoing path digests, the administrator validates its causally related components.

\section{Graph Accumulator}
\label{sec:accumulator}
This section introduces a graph accumulator that provides single-node proofs on versioned provenance graphs. A variant supporting range proofs is detailed in Appendix~\ref{ap:temporal_range_proof}.

\subsection{Hierarchical Tree Accumulation Structure}
\begin{figure}[ht]
    \centering
    \includegraphics[width=0.92\linewidth]{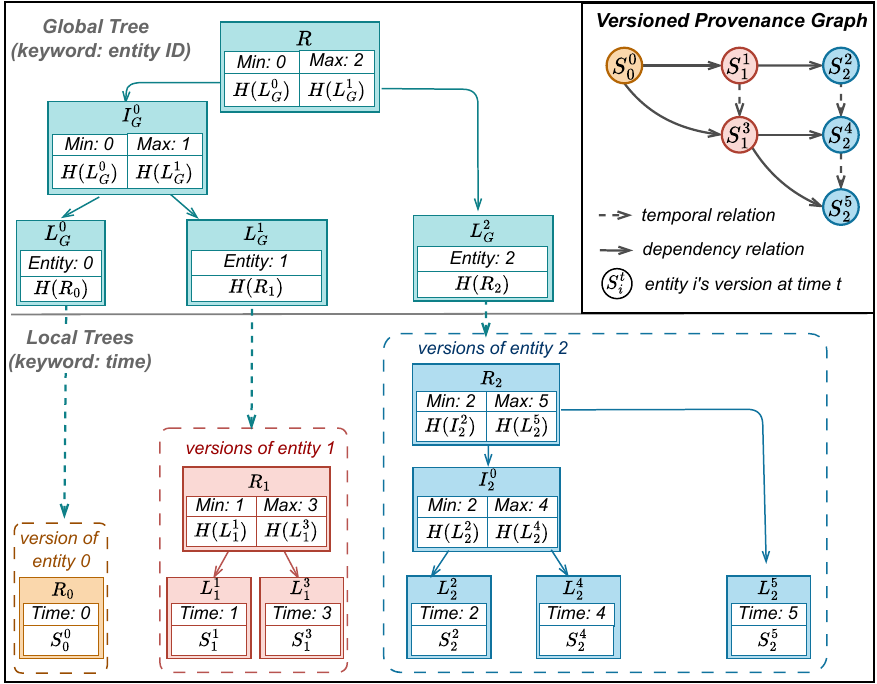}
    \caption{Graph accumulator. Here, $H(\cdot)$ is a hash function.}
    \label{fig:hierarchical_accumlator}
\end{figure}

 Our graph accumulator applies indexed Merkle trees for node storage and extends them into a hierarchical structure to support node queries in causality analysis, which involve two keywords: system entity ID and timestamp (Definition~\ref{def:causality_analysis}). The hierarchical structure is shown in Figure \ref{fig:hierarchical_accumlator}. Locally, multiple Merkle trees are used to accumulate version nodes for each system entity separately. These trees store the nodes at leaf positions and index them by timestamps, enabling proofs for timestamp-based queries. Globally, a single Merkle tree aggregates all local trees. It stores the roots of the local trees at leaf positions and indexes them by entity IDs, providing proofs for entity-based queries. By combining the proofs from both levels, our accumulator can produce complete proofs for any node queries. Notably, the root of the global tree serves as a cryptographic commitment to the entire graph, allowing anyone with the proofs to validate the results of node queries.

In the tree structure, each internal node includes index information that records the minimum and maximum keywords in its subtree, enabling efficient binary search from the root to locate the queried node and generate the corresponding proof.

\subsection{Single-Node Proof and Validation}
\label{sec:single_proof}
Node queries in causality analysis are \emph{fuzzy}, rather than based on exact keyword matching. In particular, the timestamp of the queried node need not exactly match a given value $t$; instead, it only needs to be the closest to $t$ (Definition~\ref{def:causality_analysis}). Our accumulator supports such time-fuzzy node queries in the form $N(s, r)$, where $s$ denotes the queried system entity ID and $r$ specifies a temporal relation. We consider two types of temporal relation queries: \first~$\succeq t$, which returns the nearest timestamp greater than or equal to $t$, and \second~$\preceq t$, which returns the nearest timestamp less than or equal to $t$.

We now present the membership and non-membership proofs provided by our accumulator, which respectively confirm the existence and absence of a node matching a query.

\para{Membership Proofs.} For a node query $N(s, r)$, the membership proof is generated through a two-step search process in hierarchical Merkle trees. Specifically, it is structured as a tuple $(\rho_G, \rho_L)$, where the proof $\rho_G$ is generated by searching the global tree for the entity keyword $s$, and $\rho_L$ is generated by searching the corresponding local tree according to the temporal relation $r$. The full algorithm is detailed in Appendix~\ref{ap:single_node_proof}. Both $\rho_G$ and $\rho_L$ contain nodes along their respective search paths, enabling the verifier to reconstruct the paths and validate if the queried node satisfies the query. 

\begin{figure}[ht]
    \centering
    \includegraphics[width=3.1in]{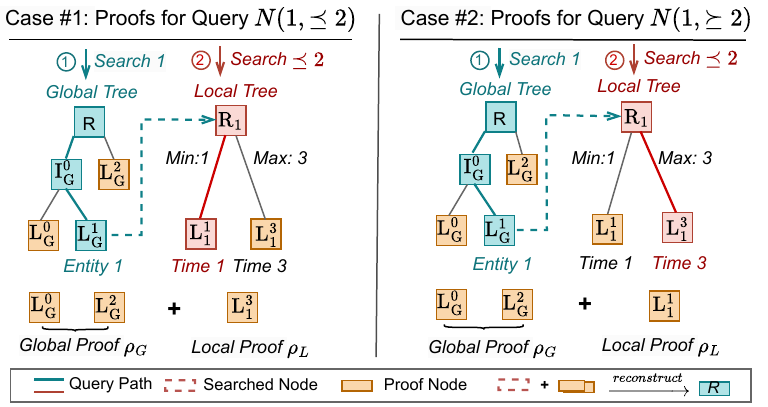}
    \caption{Single-node membership proof}
    \label{fig:node_proof}
\end{figure}

Figure \ref{fig:node_proof} shows the membership proof $(\rho_G, \rho_L)$ for query $Q(1, \preceq 2)$ (Case \#1). To generate $\rho_G$, an exact search is conducted in the global tree for the entity keyword $1$. The sibling nodes encountered along this search path constitute the proof $\rho_G$. Then, a fuzzy search is conducted in the corresponding local tree for the relation $\preceq 2$, and the sibling nodes encountered form the local proof $\rho_L$.  The figure also illustrates the proofs for query $Q(1, \succeq 2)$. Although the keywords are the same as the previous query, the different temporal relation yields a different matched node and local proof $\rho_L$.

\para{Non-membership Proofs.} Our accumulator also supports the non-membership proof for a node query, demonstrating that no nodes satisfy the query and preventing the cloud from falsely returning empty results. Similar to membership proof, it is generated through a two-step search process in hierarchical Merkle trees.
Details are provided in Appendix~\ref{ap:single_node_proof}.

\para{Validation.}
To verify (non-)membership for a node query $N(s, r)$, the verifier first reconstructs the original search path and corresponding tree structure using the provided proofs. If the root of the reconstructed tree matches the previously committed root, the proof is considered valid. The verifier then examines the index information of the reconstructed tree to validate node (non-)membership. For membership validation, if the queried node is located within the local tree of entity $s$ and satisfies the queried temporal relation $r$, membership is confirmed. For non-membership validation, if no such node is found, non-membership is confirmed.

\subsection{Dynamic Indexed Merkle Tree}
\label{sec:dynamic_merkle_tree}
To extend our accumulator to support node insertions and updates, we develop a dynamic indexed Merkle tree structure, denoted as \dtree. Compared to existing dynamic Merkle trees (e.g., Merkle B+ tree~\cite{smith2020dynamic}), our \dtree offers more efficient node insertion operations, achieving constant amortized computational cost. This efficiency is beneficial for system-wide versioned provenance graphs, where nodes are continuously created to encode system events.

\begin{figure}[ht]
    \centering
    \includegraphics[width=3in]{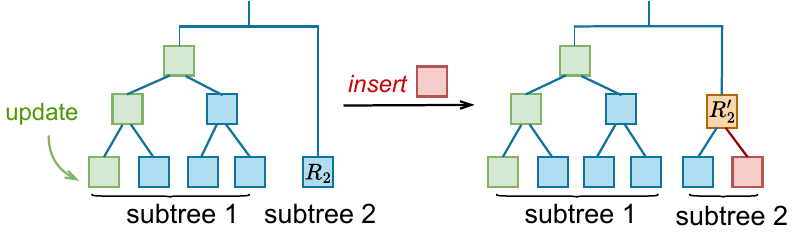}
    \caption{Structure of \dtree. Node insertion triggers a merge with the last subtree, forming a larger subtree of size $2$.}
    \label{fig:dynamic_MK_tree}
\end{figure}

\noindent\textbf{Tree Structure.} \dtree is designed to efficiently store an unbounded stream of graph nodes. To facilitate system entity-based or timestamp-based queries, \dtree indexes the nodes by entity or timestamp keywords. As the keywords typically increase over time, node insertions are arranged in temporal order, \ie each new node is appended to the current leaf node sequence. Notably, each insertion alters the tree structure and may require updating the hashes of numerous affected nodes. To improve efficiency, our \dtree adopts a multi-subtree design composed of multiple \emph{perfect binary subtrees}, as shown in Figure~\ref{fig:dynamic_MK_tree}. This design ensures that inserting a node only involves merging it with partial subtrees and updating the necessary internal hashes.

\noindent\textbf{Insertion.} Node insertion in \dtree is essentially a subtree merge process. Each new node is treated as a height-1 subtree and recursively merged with preceding subtrees from right to left, forming a larger perfect binary subtree. The detailed algorithm is provided in Appendix~\ref{ap:node_insertion_algorithm}. Compared to conventional Merkle trees, \dtree incurs lower insertion overhead, as only the hashes of affected internal nodes in partial subtrees need updating. Specifically, the amortized time complexity per insertion is $O(1)$. For instance, inserting $2^N$ nodes requires only $2^N$ subtree merges and hash updates, resulting in a near-constant average insertion time.

\para{Update.} When a node changes, the hashes of internal nodes along the path to the root of the corresponding subtree are recomputed. The update cost is $O(\log N_{sb})$, similar to the $O(\log N)$ in
standard Merkle trees, but more efficient in practice, as the subtree size $N_{sb}$ is smaller than the total size $N$.

After a series of insertions and updates, the entire \dtree should be \emph{finalized} by recursively merging the subtree roots from right to left to produce the overall tree root.
When multiple \dtrees are organized hierarchically (\eg in a graph accumulator), local trees should be finalized first, followed by the global tree, to maintain consistency and efficiency.

\section{Verifiable Versioned Provenance Graph}

\subsection{Incoming Path Digests}
\label{sec:incoming_path_digests}
According to Definition \ref{def:causality_analysis}, a node's backward causally related components lie along its incoming paths. To generate proofs for them, we employ an \emph{incoming path hash mechanism} to compute an incoming path digest $\Pi_I$ per node, representing the digest of all components in its incoming paths.

\para{Incoming Path Hash.} The hash mechanism begins by computing incoming path digests from entry nodes. As an entry node has no incoming edges, its incoming path digest $\Pi_I$ is initialized as $H(\varnothing)$, where $H(\cdot)$ denotes a hash function over a set. The incoming path digests of subsequent nodes are then derived based on their predecessors. Specifically, for node $n$, its incoming path digest is defined as follows:
\begin{equation}
\label{eq:incoming_path_digest}
n.\Pi_I = H(\{( \underbrace{s(e)}_{\text{in-edge}}||\underbrace{e.src.\Pi_I}_{\text{a predecessor node's $\Pi_I$}}) \mid e \in n.E_{in}\})
\end{equation}
where $n.E_{in}$ denotes the incoming edges of $n$, and $s(e)$ is the string representation of an edge, formed by concatenating the associated nodes’ fields \texttt{[id $\mathbin{||}$ time $\mathbin{||}$ \ldots]} and event data \texttt{[event\_type $\mathbin{||}$ \ldots]}. The symbol $\mathbin{||}$ denotes concatenation.

\para{Dynamic Generation of ${\Pi_I}$.} In a versioned provenance graph, nodes and edges are created continuously to record system events (see \S\,\ref{sec:provenance_background}). Upon the creation of a new node, its incoming path digest $\Pi_I$ should be dynamically computed based on the previously connected nodes and edges, as defined in Equation~\ref{eq:incoming_path_digest}. Once generated, the digest remains unchanged. This immutability arises from the properties of versioned provenance graphs: each node represents a version of an entity, and its incoming edges capture its creation histories. Since this history is immutable, the node cannot receive new incoming edges, and its incoming path digest remains unchanged.

\subsection{Backward Causality Relation Validation}
\label{sec:backward_validation}
\para{Proof.} According to Equation~\ref{eq:incoming_path_digest}, a node’s incoming path digest is computed via a recursive hash mechanism, capturing all components along its incoming paths, \ie backward causally related components. Thus, the incoming path digest of the node and its backward-connected components can serve as a cumulative proof for its backward causality relations.

\para{Validation.} To verify a node’s backward causality relations, we recursively hash the components along its incoming paths from entry nodes to \emph{regenerate} an incoming path digest. If the regenerated digest matches the node's original one, the integrity of the node’s backward causality relations is confirmed. The detailed algorithm is described in Algorithm~\ref{alg:backward_verification}.

\subsection{Outgoing Path Digests}
\label{sec:outgoing_path_digests}
A node’s forward causally related components lie along its outgoing paths. To generate proofs for them, we employ an \emph{outgoing path hash mechanism} that computes an outgoing path digest $\Pi_O$ per node, capturing all related components.
 
\para{Outgoing Path Hash.} The hash mechanism operates inversely to incoming path hash: it starts from exit nodes (\ie nodes without outgoing edges). For an exit node, its outgoing path digest $\Pi_O$ is computed as $H(\varnothing)$, where $H(\cdot)$ represents a hash function. Tracing backward, the digest of each preceding node is derived from the digests of its successors. Specifically, for node $n$, its outgoing path digest is defined as follows:
\begin{equation}
\label{eq:outgoing_path_hash}
        n.\Pi_{O} = H(\{(\underbrace{s(e)}_{\text{out-edge}}~||~\underbrace{e.dst.\Pi_O}_{\text{a successor node's $\Pi_O$}}) \mid e \in n.E_{out}\})
\end{equation}
where $n.E_{out}$ denotes $n$'s outgoing edges,  and $s(e)$ is the string representation of edge $e$ formed by concatenating related node fields and event data. The symbol $||$ denotes concatenation. 

Notably, a node’s $\Pi_O$ changes when new outgoing edges are added. To update efficiently, we adopt ordering-invariant incremental hashing~\cite{bellare1997new} over a set, enabling homomorphic addition (or subtraction) of component digests to the node’s $\Pi_O$ without recomputing the entire digest.

\para{Dynamic Generation of ${\Pi_O}$.} In a versioned provenance graph, nodes are continuously added to encode system events. Hence, a node’s outgoing path digest ${\Pi_O}$ should be generated dynamically as events occur. Since a newly created node has no outgoing edges, its ${\Pi_O}$ is initialized as $H(\varnothing)$.

\begin{figure}[ht]
    \centering
    \includegraphics[width=3.1in]{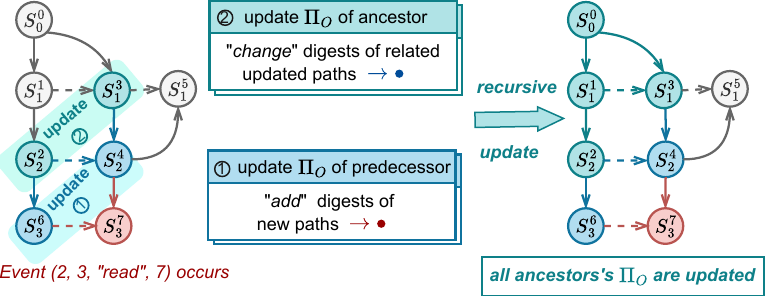}
    \caption{Updating outgoing path digests upon event occurrence. Here, $\Pi_O$ denotes outgoing path digests}
    \label{fig:outgoing_path_digest}
\end{figure}

\para{Dynamic Update of ${\Pi_O}$.} Along with node creation, the graph also generates associated temporal and dependency edges. These components alter the outgoing path structures of connected ancestors, necessitating updates to their $\Pi_O$. As shown in Figure \ref{fig:outgoing_path_digest}, event $(2, 3, \text{"read"}, 7)$ triggers the creation of node $S_3^7$ along with edges $e_7^t$ and $e_7^4$. These components affect the outgoing path structures of nodes $\{S_0^0, S_1^1, S_1^3, S_2^2, S_2^4, S_3^6\}$, requiring corresponding updates to their $\Pi_O$. 

The update process is recursive, beginning with the immediate predecessors of the newly created node. 

\phreset
\phitem{Update $\Pi_O$ of Immediate Predecessors:} The addition of $S_i^j$ and its associated edges expands the outgoing path structures of their predecessors. To synchronize the changes, we \emph{add} the digest of the new elements to  each predecessor $p$'s original $\Pi_O$ using an incremental hash function, as follows: 
    \begin{equation}
    \label{eq:update_immediate_pre}
    p.\Pi_{O} ~\oplus= H(s(e_{p \rightarrow S_i^j})~||~S_i^j.\Pi_{O}), \text{when $S_i^j$ is added.}
\end{equation}
Here, $e_{p \rightarrow S_i^j}$ denotes the edge from $p$ to $S_i^j$, $s(e_{p \rightarrow S_i^j})$ represents the string representation of the edge and associated nodes, and $\oplus$ signifies homomorphic hash addition.
\phitem{Update $\Pi_O$ of Earlier Ancestors:}
Updates to the immediate predecessors’ $\Pi_O$ recursively propagate to earlier ancestors. Specifically, when a node $n$ is updated,  $\Pi_O$ of its predecessors should be updated accordingly. Since $n$ typically affects only part of its predecessors’ outgoing path structures rather than the entire structure, we employ an incremental hash function~\cite{bellare1997new} to update the affected digests efficiently.

The update for each predecessor $p$ of node $n$ is defined as:
\begin{equation}
\begin{split}   
\label{eq:update_previous_ancestors}
    p.\Pi_{O} & ~\underbrace{\ominus= H(s(e_{p \rightarrow n})~||~n.\Pi_{O})}_{\text{remove old path digest related to $n$}} \\ & \oplus \underbrace{H(s(e_{p \rightarrow n})~||~n.\Pi_{O}^*)}_{\text{add new path digest related to $n$}}, \text{when $n$ is modified.}
\end{split}
\end{equation}
Here, $e_{p \rightarrow n}$ is the edge from $p$ to $n$, $s(e_{p \rightarrow n})$ denotes the string representation of the edge and connected nodes, $n.\Pi_{O}$ and $n.\Pi_{O}^*$ denotes  $n$'s old and new outgoing path digests, and $\ominus$ and $\oplus$ denote homomorphic hash subtraction and addition.

\para{Challenge of Exponential Digest Updates.} In a versioned provenance graph, each node connects to two predecessor nodes via temporal and dependency edges. Thus, if paths of length $L$ exist, adding a node may affect the outgoing path structures of $2^L$ ancestors,  
thereby resulting in exponential digest update overhead. To address this, we adopt the following update-efficient outgoing path digests.

\subsection{Segmented Outgoing Path Digests}
\label{sec:seg_outgoing_path_digests}
To reduce the update overhead of outgoing path digests, we apply the following segmentation strategies to \emph{split} specific paths at first. The strategies reduce the number and length of paths leading to a node, ensuring that node insertion affects fewer ancestors, thereby reducing digest update overhead. 

\begin{figure}[ht]
    \centering
    \includegraphics[width=3.2in]{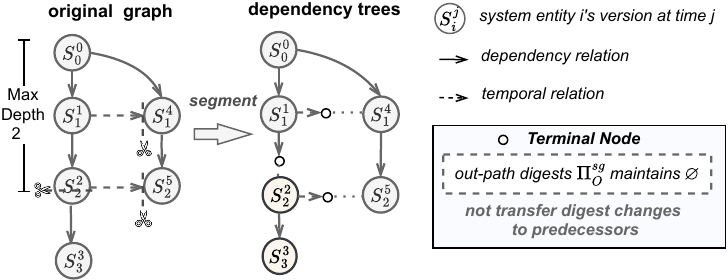}
    \caption{Segmentation strategies}
\label{fig:segmentation}
\end{figure}

\phreset
    \phitem{Temporal Edge Segmentation:} This strategy detaches each temporal edge from its destination node. The original destination is replaced with a terminal node. The terminal node has no outgoing edges and never receives new ones, so its outgoing path digests remain empty. To preserve logical connectivity, it maintains a pointer to the original destination. As shown in Figure \ref{fig:segmentation}, this segmentation converts the provenance graph into multiple trees, where each node has only one incoming dependency edge. Consequently, when a node is created, only its ancestors along a single path require updates, reducing the update overhead to linear complexity.
    
    \phitem{Deep Tree Segmentation:}  While the above strategy reduces digest update complexity by forming smaller trees, updates still become time-consuming if trees grow too deep. To address this,  we define a segmentation depth $L$. If a dependency edge causes the dependency depth of a tree to exceed $L$, the edge and connected nodes are moved into a new tree. 

After segmentation, each node’s outgoing path digest over the segmented paths ($\Pi_O^{sg}$) can be computed using Equation~\ref{eq:outgoing_path_hash}. Similar to the original $\Pi_O$, a node's $\Pi_O^{sg}$ is initialized as $H(\varnothing)$ when a new event triggers its creation. 

\begin{figure}[ht]
    \centering
    \includegraphics[width=3.1in]{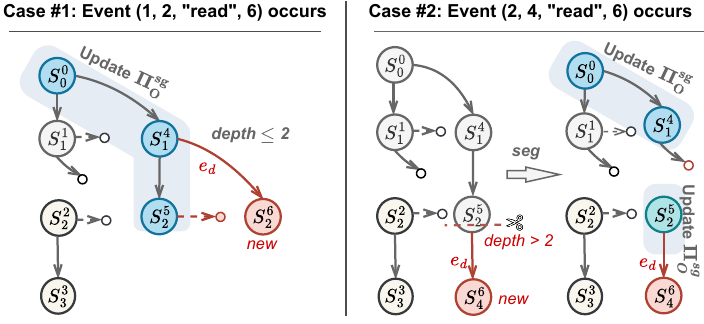}
    \caption{Updating segmented outgoing path digests $\Pi_O^{sg}$ upon event occurrence. Here, the tree segmentation depth is 2.}
    \label{fig:segmented_out_path_digest}
\end{figure}

\noindent\textbf{Dynamic Update of ${\Pi_O^{sg}}$}. As with the original outgoing path digests, $\Pi_O^{sg}$ should be dynamically updated when associated dependency and temporal edges are added to the graph. However, the update process is more complex, as edge additions may trigger different structural changes and corresponding update procedures. According to segmentation strategies, the temporal edge does not increase the tree's dependency depth and thus remains within the original tree. Thus, affected ancestors’ $\Pi_O^{sg}$ can be directly updated using Equations~\ref{eq:update_immediate_pre} and~\ref{eq:update_previous_ancestors}. In contrast, the dependency edge increases the dependency depth of the tree. As shown in Figure \ref{fig:segmented_out_path_digest}, this may result in two structural changes, each requiring a distinct update:

\phreset
\phitem{Case \#1.} If the dependency edge $e^d$ does not increase the tree depth beyond the threshold $L$, the edge and its connected nodes remain in the tree. In this case, $\Pi_O^{sg}$ of connected ancestors can be updated using Equations \ref{eq:update_immediate_pre} and \ref{eq:update_previous_ancestors}.

\phitem{Case \#2.} If the dependency edge $e^d$ increases the tree depth beyond the threshold $L$, the edge is moved to a new tree. The connected nodes $S_2^5$ and $S_4^6$ are also transferred to the new tree, and the original position of $S_2^5$ in the old tree is replaced with a terminal node.  Structural changes in the old and new trees affect different nodes. In the original tree, the added terminal node influences its ancestors $S_0^0$ and $S_1^4$, while in the new tree, the new edge $e^d$ affects only root $S_2^5$. To accommodate the changes, we update $\Pi_O^{sg}$ of the affected nodes in the old and new trees using Equations \ref{eq:update_immediate_pre} and~\ref{eq:update_previous_ancestors}, respectively.

In summary, compared to original outgoing path digests with exponential update overhead, the use of segmented outgoing path digests reduces this overhead to linear complexity $O(L)$, where $L$ is the predefined tree segmentation depth.

\subsection{Forward Causality Relation Validation}
\label{sec:forward_validation}
\para{Proof.} A node’s outgoing path digest $\Pi_O$ represents the hash of all components along its outgoing paths (\ie forward causally related components). Thus, $\Pi_O$ of a node and its forward-connected nodes can serve as a cumulative proof of its forward causality relations. In practice, however, the proof process may be more complex, as segmented outgoing path digests $\Pi_O^{sg}$ are often used instead of $\Pi_O$ for efficiency. 

A node's segmented outgoing path digest $\Pi_O^{sg}$ captures only the components within a single dependency tree and does not encompass all forward causally connected components. Therefore, utilizing the node's $\Pi_O^{sg}$ is insufficient for validating its complete causality relations. To ensure full validation,  $\Pi_O^{sg}$ of roots from subsequent connected trees should be utilized to validate the remaining components. Notably, to confirm the authenticity of the $\Pi_O^{sg}$, node proofs of the roots should also be retrieved from the graph accumulator.

\begin{figure}[ht]
    \centering
    \includegraphics[width=3.3in]{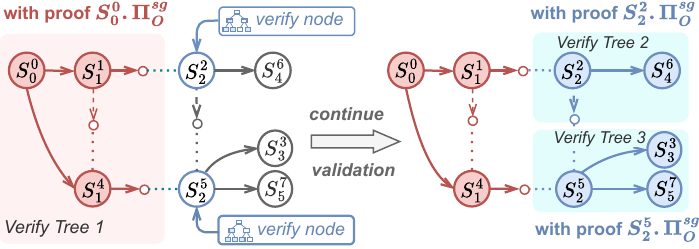}

    \caption{Forward relation validation with $\Pi_O^{sg}$}

    \label{fig:seg_forward_validation}
\end{figure}

\para{Validation.} Figure \ref{fig:seg_forward_validation} illustrates the validation process using segmented outgoing path digests $\Pi^{sg}_O$. The process begins by using the initial node’s $\Pi^{\textit{sg}}_O$ to validate forward causally related components within the dependency tree $1$. If successful, the roots of subsequent trees (\ie $S_2^2$ and $S_2^5$) are then validated using node proofs from the graph accumulator. To enhance the efficiency of this process, temporal range proofs can be applied to enable batch validation for these roots (see Appendix~\ref{ap:temporal_range_proof}). Once the roots are validated, their $\Pi^{sg}_O$ are then used to verify the remaining connected components. 
The detailed validation algorithm is given in Algorithm~\ref{alg:forward_verification}.

\section{Verifiable Causality Analysis System}
\label{sec:vCause}
By integrating our graph accumulators and versioned provenance graphs, \sys provides complete proofs for causality analysis. The system involves three parties---endpoint loggers, a cloud, and a security administrator---and supports four core operations: \emph{event recording}, \emph{commitment}, \emph{causality analysis proof}, and \emph{validation}, as described below:

\begin{figure}[t]
    \centering
    \includegraphics[width=\linewidth]{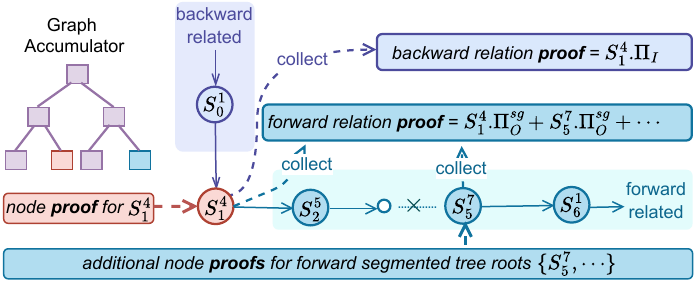}
    \caption{Proof for a causality analysis query. The query targets entity $1$ at timestamp $t_4$. $S_1^4$ is the corresponding node; $\Pi_I$ and $\Pi_O^{sg}$ denote incoming/segmented outgoing path digests.}
    \label{fig:sys_proof}
\end{figure}
\para{Event Recording.} When a system event occurs, the endpoint logger updates its local verifiable versioned provenance graph by creating a new node representing the current system entity version, along with its associated temporal and dependency edges. It then computes the node’s incoming and (segmented) outgoing path digests and updates the digests of affected ancestors (see \S\,\ref{sec:incoming_path_digests} and \S\,\ref{sec:seg_outgoing_path_digests}).

\para{Commitment.}  To enable timely validation, each endpoint logger periodically commits node changes during event recording into the Merkle trees of a graph accumulator~(\S\,\ref{sec:dynamic_merkle_tree}). It then signs the tree root $R$ along with the current timestamp $t$ to produce a fresh signature $\sigma_R$. The tuple $(R, t, \sigma_R)$ serves as the cryptographic commitment to the entire graph and is sent to the cloud along with the corresponding event logs.

\para{Causality Analysis Proof.}
To generate proofs for causality analysis queries, the cloud reconstructs a verifiable versioned provenance graph and a graph accumulator from the logs of each endpoint. As shown in Figure~\ref{fig:sys_proof}, upon receiving a query for an endpoint's system entity at a given timestamp, the cloud first locates the corresponding graph node (\ie $S_1^4$) and its causally related components. It then uses the accumulator to generate a proof for the node (\S\,\ref{sec:single_proof}) and collects its incoming and outgoing path digests as evidence of backward and forward causal relationships (\S\,\ref{sec:backward_validation} and \S\,\ref{sec:forward_validation}). If segmented outgoing digests are applied, the forward causally related components may be segmented accordingly. For completeness, the cloud also needs to retrieve node proofs of the corresponding segmented tree roots from the accumulator.

\para{Causality Analysis Validation.} 
To validate causality analysis results, the administrator retrieves the latest cryptographic graph commitment from the cloud and verifies its signature. If valid, the administrator validates the queried node using its node proof~(\S\,\ref{sec:single_proof}), and then validates backward and forward causally related components with the node’s incoming and (segmented) outgoing path digests~(\S\,\ref{sec:backward_validation} and \S\,\ref{sec:forward_validation}).
\section{Security Analysis}
\label{sec:security}
We follow the notion of existential unforgeability under chosen message attacks (EUF-CMA)~\cite{dodis2012message} to define the security of \sys against an adaptive adversary $\mathcal{A}$ that can add, delete, or modify components in causality analysis results. The property is formalized via the following game $Forge_{\mathcal{A}}^{\sys}(k)$:

\phreset
    
    \phitem{Causality Analysis Challenges.} $\mathcal{A}$ makes a polynomial number of queries $Q$ to the causality analysis oracle. For each query $q \in Q$, the oracle identifies a point-of-interest node $n$ and returns its backward and forward causally related components: $\{V_{\rightarrow n}, E_{\rightarrow n}\}$ and $\{V_{n \rightarrow}, E_{n \rightarrow}\}$. Additionally, it provides the corresponding commitment $(R, t, \sigma_R)$ and the node proof $\rho_n$ for $n$. The internal incoming and outgoing path digests of $n$ are treated as its causality analysis proofs.
    
    \phitem{Forgery.} $\mathcal{A}$ outputs a forged causality analysis result $C^* = \{n^*, \{V_{\rightarrow n}^*, E_{\rightarrow n}^*\}, \{V_{n \rightarrow}^*, E_{n \rightarrow}^*\}\}$ for a query $q^* \notin Q$, along with a forged proof $\rho_n^*$ and a commitment $(R^*, t^*, \sigma_R^*)$.
    
    \phitem{Verification.} $\mathcal{A}$ submits the forged results $C^*$, proof $\rho_{n}^*$, and commitment $(R^*, t^*, \sigma_R^*)$. The challenger first validates the signature of the commitment. It then checks whether $C^*$ and $\rho_{n}^*$ are valid with respect to $q^*$. If all verifications pass, the adversary wins the game and outputs 1.

\begin{myDef}
{\rm \textbf{(Unforgeability of Causality Analysis Results)}} We say
\sys achieves unforgeability of causality analysis results if, for any polynomial-time adversary $\mathcal{A}$:
\begin{equation}
    \Pr(Forge_{\mathcal{A}}^{\sys}(k) = 1) \leq \text{negl}(k)
\end{equation}
where $\text{negl}(k)$ denotes a negligible function.
\end{myDef}

Now, we give the following security theorem. (formal security theorem and proof can be found in Appendix~\ref{ap:sec_analysis})
\begin{myTheo}
\label{theo:vCause_sec}
{\rm \sys} achieves unforgeability of causality analysis results if the signature scheme is EUF-CMA-secure~\cite{dodis2012message}, the Merkle tree structure is position-binding~\cite{catalano2013vector}, and the verifiable versioned provenance graph upholds causality relation unforgeability~(Appendix~\ref{ap:graph_security}).
\end{myTheo}

\section{Experimental Evaluation}
\para{Implementation.} 
We implement a prototype of \sys in 3,500 lines of C++ code using OpenSSL 1.1.1w\cite{OpenSSL}. The graph accumulator adopts a hierarchical indexed Merkle tree built with SHA3-256 hash function, while digests in the versioned provenance graph leverage an incremental hash function~\cite{bellare1997new}.  To generate a cryptographic commitment for the graph, the Merkle root is signed using ECDSA~\cite{johnson2001elliptic}.

\begin{table}[ht]
\small
\caption{Summary of datasets.  $|V|$ and $|E|$ denote the number of versioned nodes and edges in versioned provenance graphs built from a dataset. SS. denotes the StreamSpot dataset.}
\label{tb:dataset}
\begin{tabular}{p{0.4in}<{\centering}p{0.6in}<{\centering}p{0.5in}<{\centering}p{0.5in}<{\centering}p{0.5in}<{\centering}}
\toprule
\multirow{2}{*}{\textbf{Dataset}}    & \textbf{\# of Endpoints} & \textbf{Avg \# of Events} & \multirow{2}{*}{\textbf{Avg} $\bf{|V|}$} & \multirow{2}{*}{\textbf{Avg} $\bf{|E|}$}  \\ \midrule
SS. & 600          & 149K          & 149K    & 290K          \\ \midrule
OpTC & 1,000 & 17M & 17M & 32M \\ \bottomrule
\end{tabular}
\end{table}

\para{Datasets.}  We evaluate \sys on two real-world endpoint auditing datasets: DARPA OpTC~\cite{OpTC} and StreamSpot~\cite{manzoor2016fast}, both containing large volumes of realistic system event logs. Table~\ref{tb:dataset} summarizes their key statistics. The StreamSpot dataset contains system events from six simulated web browsing scenarios, each executed 100 times. On average, each execution yields thousands of event logs. In our experiments, we treat these 600 executions as logs from 600 endpoints. The OpTC dataset, produced by DARPA's Transparent Computing (TC) program, includes over 17 billion events collected from 1,000 real hosts over six days in an enterprise network.

Experiments are performed on a Ubuntu 22.04 PC with 4 Intel Xeon 3.60GHz processors and 32GB RAM. 
The evaluation covers \first efficiency of proposed authenticated data structures (\S\, \ref{sec:exp_graph_accumulator} and \S\,\ref{sec:exp_vGraph}); \second \sys's computational costs~(\S\,\ref{sec:exp_overall_cost}); \third runtime overhead under realistic deployment settings~(\S\,\ref{sec:exp_overhead}); \fourth communication and storage costs~(\S\,\ref{sec:exp_storage_comm}).

\subsection{Efficiency of Graph Accumulators}
\label{sec:exp_graph_accumulator}
Recall that the computational cost of our accumulators mainly depends on the number of accumulated nodes. To evaluate its efficiency, we use data from an endpoint in the larger dataset (OpTC), yielding a versioned provenance graph with 25 million nodes. We then measure the execution time of four operations: insertion, update, proof generation, and validation.

\begin{figure}[ht]
  \centering
  \subfloat[Total Insertion Delay]{\includegraphics[width=1.64in]{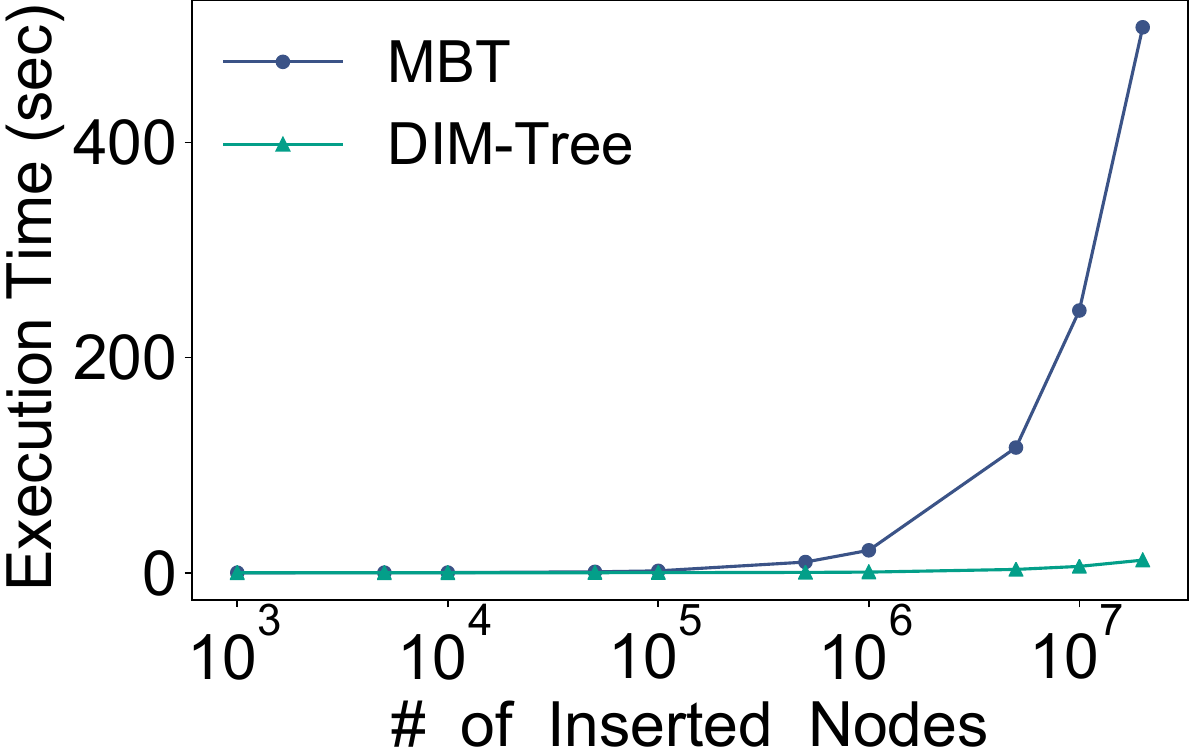}
  }\hfill
   \subfloat[Update Time]{\includegraphics[width=1.65in]{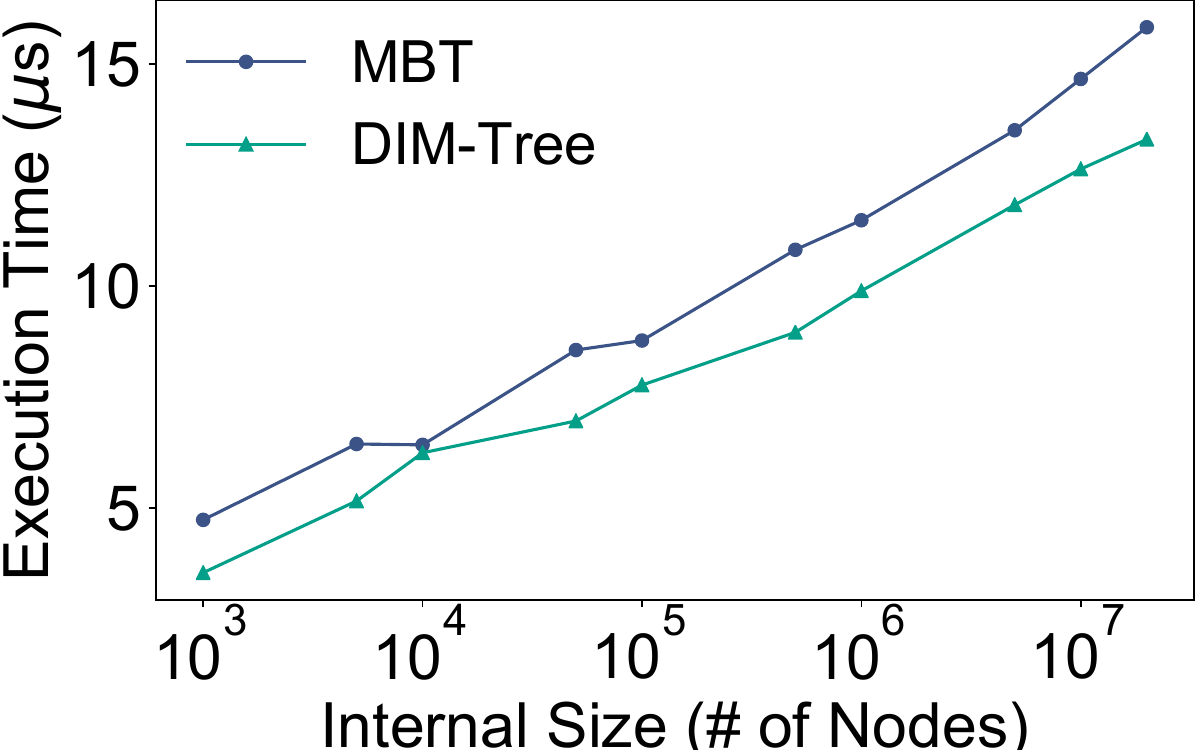}}
   \\
   \subfloat[Single-Node Proof Time]{\includegraphics[width=1.64in]{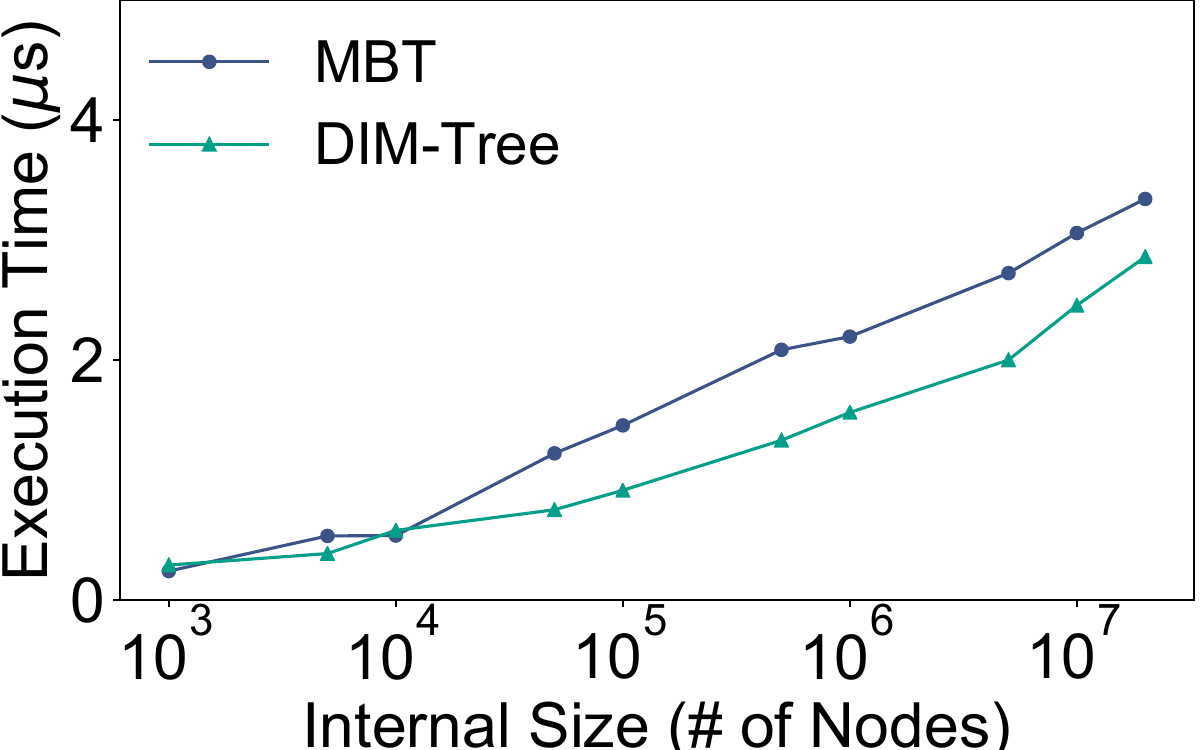}
  }\hfill
  \subfloat[Temporal Range Proof Time]{\includegraphics[width=1.64in]{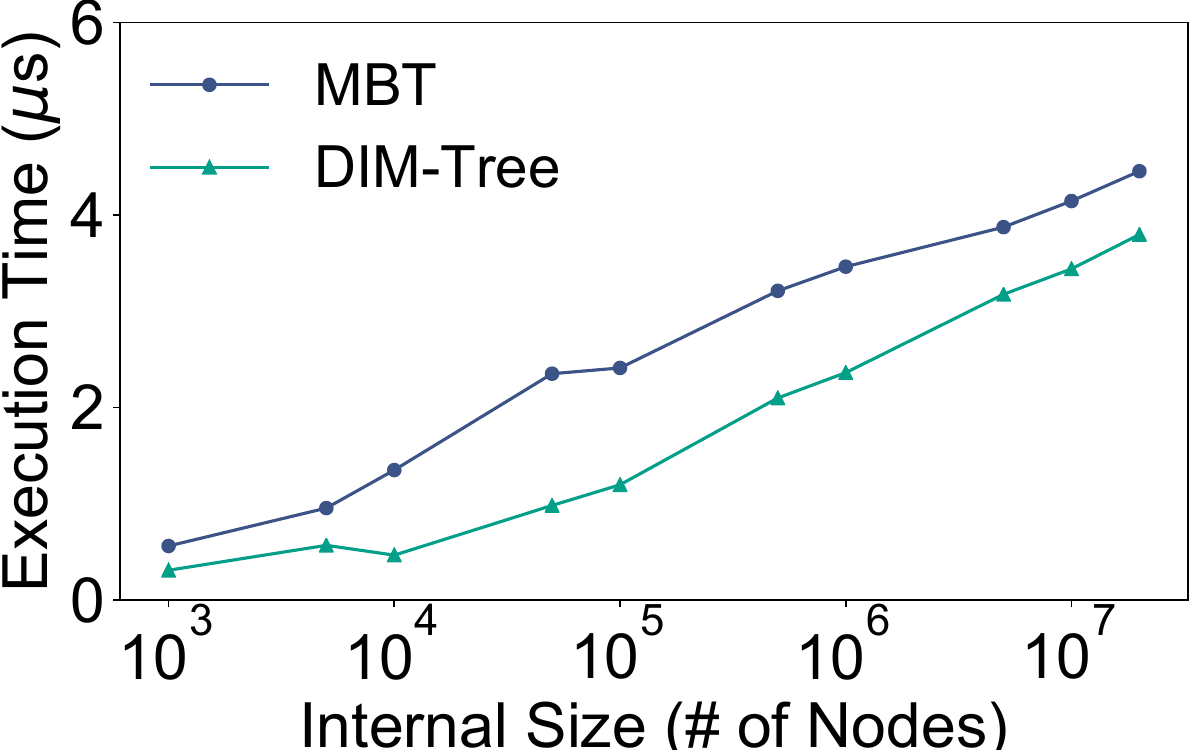}}
  \\
   \subfloat[Single-Node Validation Time]{\includegraphics[width=1.64in]{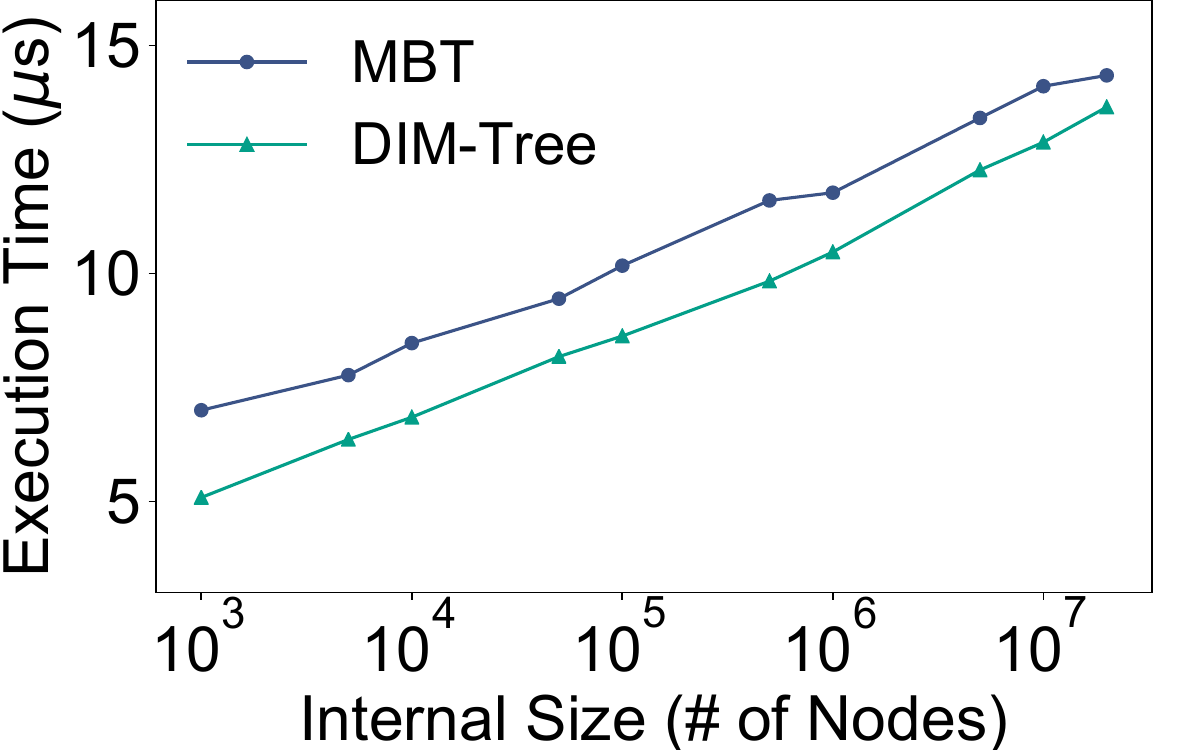}}
   \hfill
   \subfloat[Temporal Range Validation Time]
   {\includegraphics[width=1.64in]{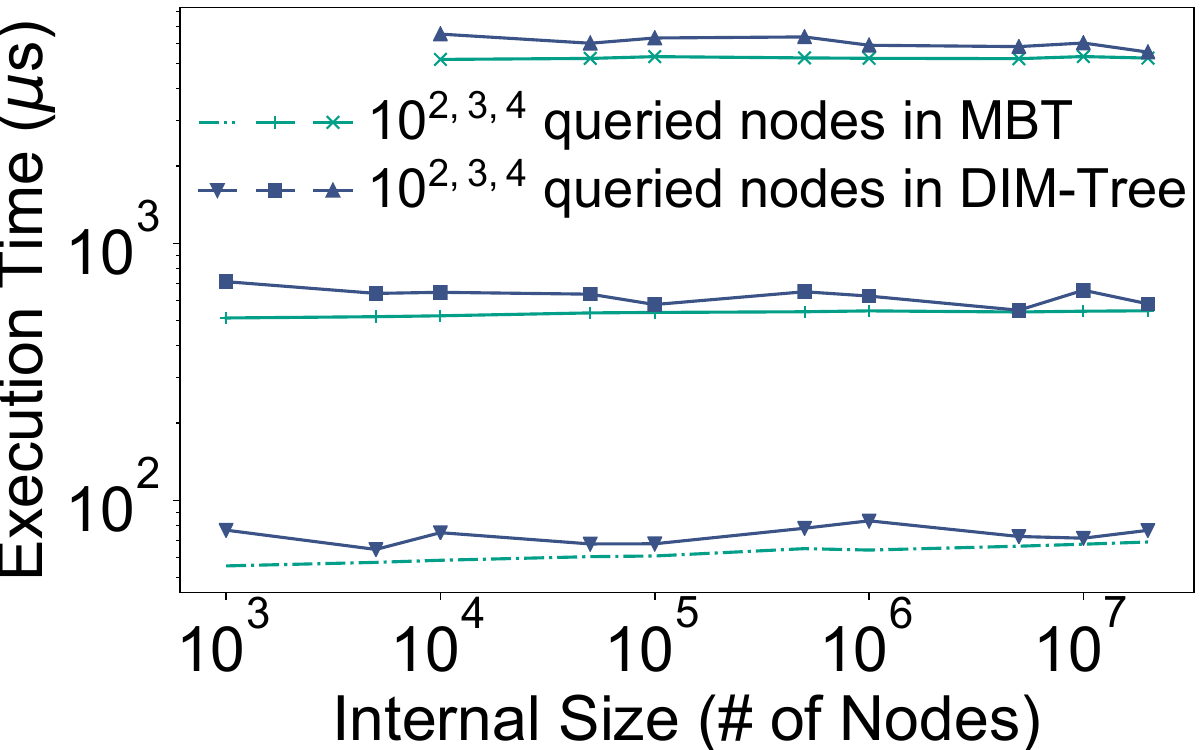}}\\
   \caption{Performance of graph accumulators. Here, MPT refers to the Merkle B+ tree-based accumulator, while \dtree denotes our \dtree-based accumulator.}
    \label{exp:graph_accumulator}
\end{figure} 

Notably, our accumulator adopts a tailored dynamic Merkle tree, \dtree, which supports efficient node insertion with $O(1)$ amortized time complexity. To evaluate its efficiency, we implement a baseline accumulator using a conventional dynamic Merkle tree (Merkle B+ tree) and compare both designs using the same 25-million-node dataset.

\para{Node Insertion.} Figure~\ref{exp:graph_accumulator}.(a) presents the insertion time for different node counts. Compared to the Merkle B+ tree-based design, our \dtree-based accumulator demonstrates significantly faster performance. Its insertion time grows linearly with the number of nodes, confirming an average constant cost per insertion. In total, inserting 25 million nodes takes $\approx  14$ seconds, with each insertion costing about 0.57~$\mu$s.

\para{Node Update.}   The time of updating a node in the accumulator depends on the number of internal nodes. To assess this, we measure the update latency across different internal sizes. As shown in Figure~\ref{exp:graph_accumulator}.(b), the update time of both accumulators increases with the number of internal nodes. Particularly, our \dtree-based accumulator achieves slightly lower update time than the Merkle B+ tree-based accumulator.

\para{Node Proof.} To comprehensively evaluate the proof time for both single-node and temporal range queries, we vary the number of stored nodes in each accumulator and generate proofs for 1,000 single-node and temporal range queries. Figures~\ref{exp:graph_accumulator}.(c) and \ref{exp:graph_accumulator}.(d) show the average proof time for both query types. We observe that the proof time increases with the number of internal nodes.

\para{Node Validation.} 
We evaluate the validation time for both single-node and temporal range queries. Recall that validation time depends on both the size of the original tree and the number of queried nodes (see \S\,\ref{sec:single_proof} and Appendix~\ref{ap:temporal_range_proof}). To comprehensively assess this cost, we vary the internal size of each accumulator, generate proofs for 3,000 single-node and temporal range queries, and perform validation. Particularly, temporal range queries are grouped into three scales:  $10^3$, $10^4$, and $10^5$ successive nodes within a given timestamp range. Figures~\ref{exp:graph_accumulator}.(e) and~\ref{exp:graph_accumulator}.(f) show the average validation time for single-node and temporal range queries, respectively. The results indicate that single-node validation time primarily increases with the number of internal nodes, while temporal range validation time scales with the number of queried nodes.

\subsection{Performance of Verifiable Versioned Provenance Graphs}
\label{sec:exp_vGraph}
We evaluate our verifiable versioned provenance graphs by constructing two graph instances from logs in StreamSpot and OpTC, containing 716K and 25M nodes, respectively. For both graphs, we measure the generation and update time of incoming and outgoing path digests during construction.

 \begin{table}[ht]
\centering
\small
\caption{Average generation time of path digests}
\label{tab:path_digest_generation}
\begin{tabular}{p{1.6in}cp{0.6in}<{\centering}}
\toprule
\textbf{Operation}                 & \textbf{StreamSpot} & \textbf{OpTC} \\ \midrule
Incoming path digest generation & 1.59 $\mu$s         & 1.64 $\mu$s           \\ 
Outgoing path digest generation & 1.34 $\mu$s         & 1.37 $\mu$s           \\ \bottomrule
\end{tabular}
\end{table}

\para{Digest Generation.} A node's incoming and outgoing path digests are generated immediately upon its creation. Specifically, the outgoing path digest is always initialized to $H(\varnothing)$, while the incoming path digest is computed based solely on the node’s intermediate predecessors. Thus, their generation time is independent of the graph structure. Results from the StreamSpot and OpTC datasets confirm this. As shown in Table~\ref{tab:path_digest_generation}, the average generation time for both digests remains consistent across datasets, ranging from 1 to 2~$\mu$s.

\begin{figure}[ht]
  \centering
  \subfloat[Average update time of outgoing path digests]{\includegraphics[width=3.3in]{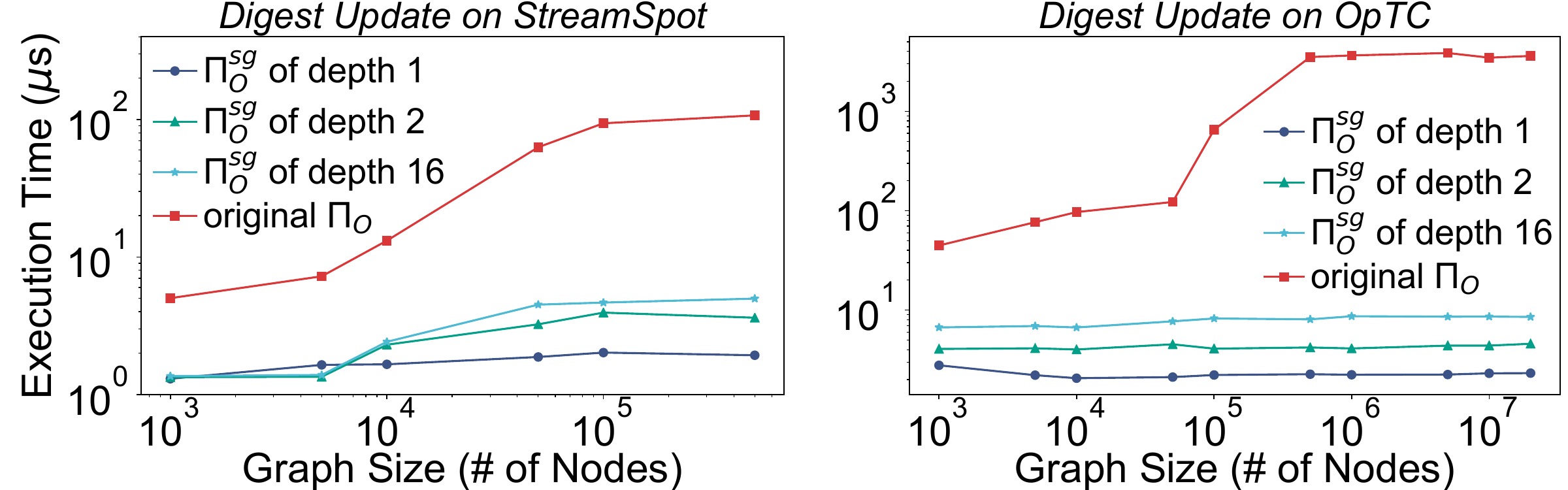}
  }\\
  \subfloat[Backward causality relation proof and validation time]{\includegraphics[width=3.3in]{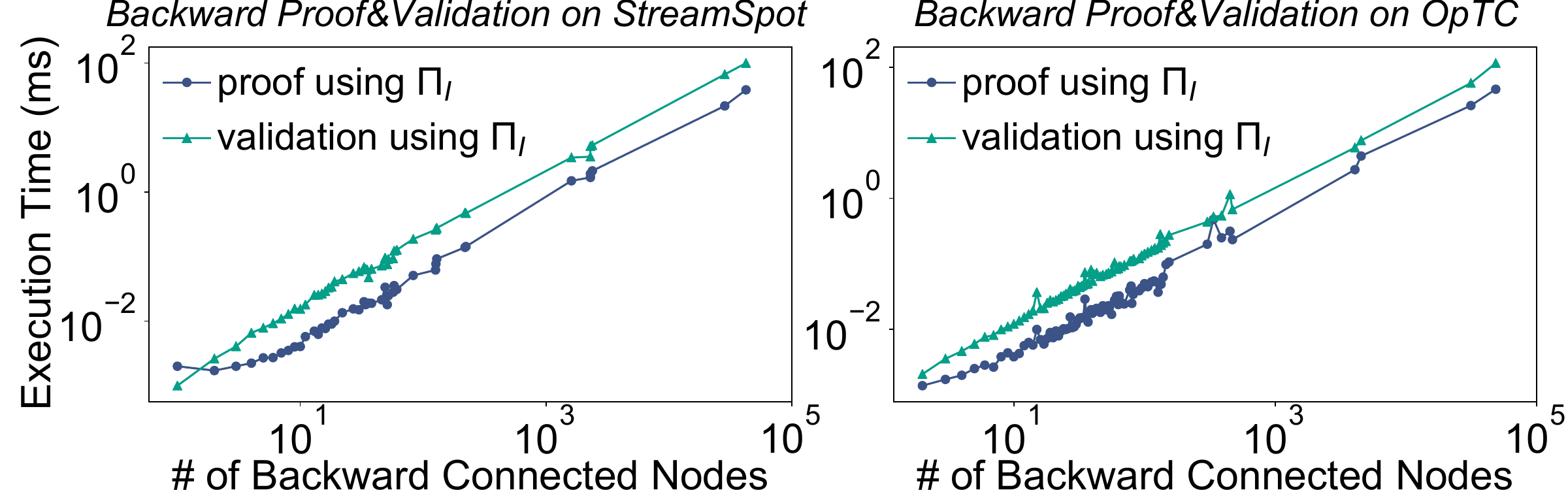}
  }\\
\subfloat[Forward causality relation proof and validation time]{\includegraphics[width=3.3in]{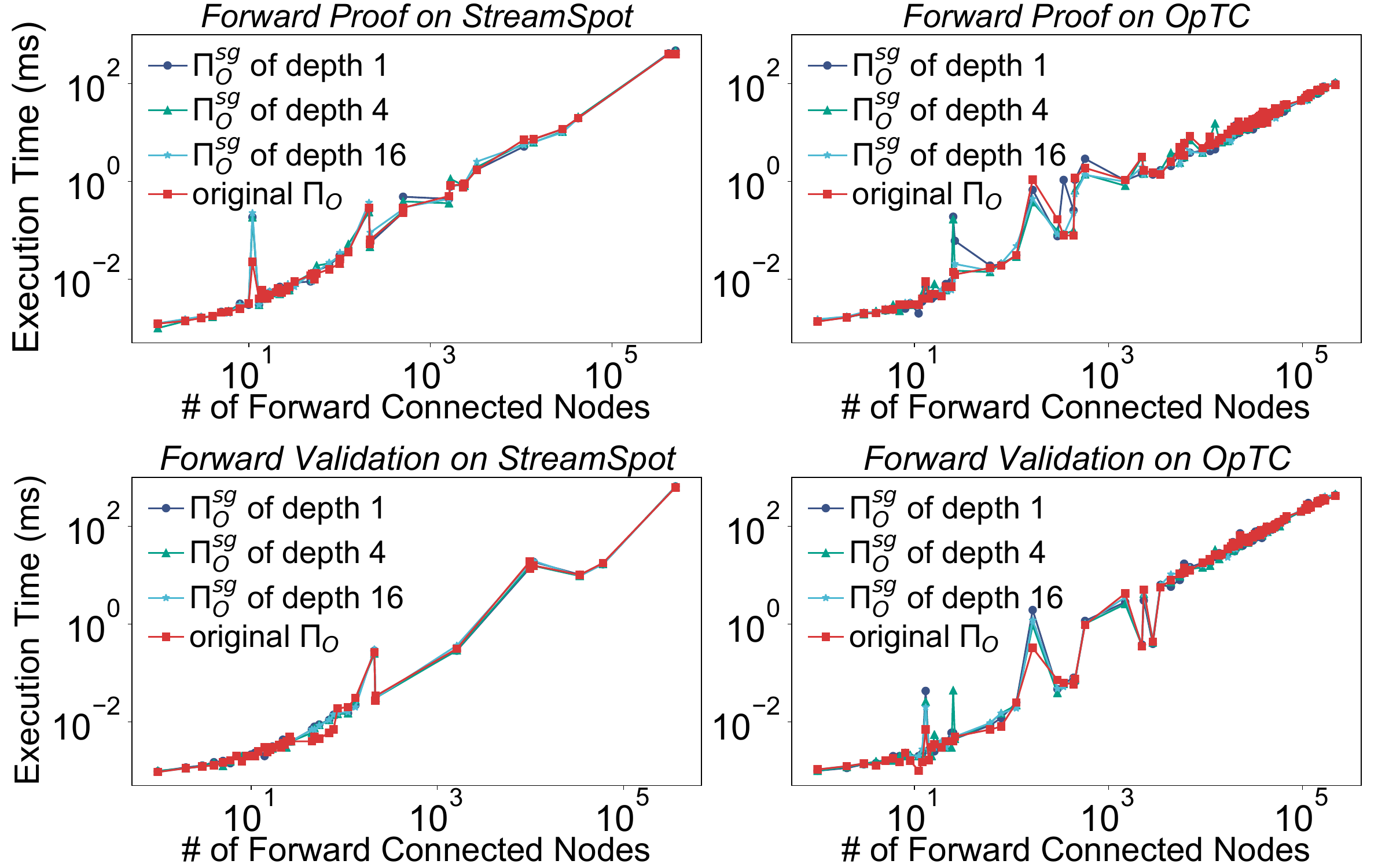}}
    \caption{Performance of verifiable versioned provenance graphs}
\label{exp:vGraph}
\end{figure}

\para{Digest Update.} We do not consider the update of incoming path digests as they remain unchanged once generated due to properties of versioned provenance graphs~(see \S \,\ref{sec:incoming_path_digests}). Our experiments focus on measuring the update time of outgoing path digests ($\Pi_O$). According to Equations \ref{eq:update_immediate_pre} and \ref{eq:update_previous_ancestors}, the update process is triggered by node insertion. The insertion changes the outgoing paths of connected ancestors, necessitating updates to their $\Pi_O$. Thus, the overall update time grows with the graph size, since larger graphs tend to have more affected ancestors. To assess this impact, we generate graphs of varying sizes from StreamSpot and OpTC datasets. Figure~\ref{exp:vGraph}.(a) shows the average update time of $\Pi_O$ across these graphs, revealing a substantial increase as graph size grows.

Additionally, we evaluate the update time of segmented outgoing path digests ($\Pi_{O}^{sg}$), which are optimized variants of the original $\Pi_O$. These digests are generated based on graph segmentation strategies, reducing the update time to linear complexity $O(L)$, where $L$ is the predefined segmentation depth~(\S\,\ref{sec:seg_outgoing_path_digests}). To assess the impact of $L$, we vary its value and measure the corresponding average update time for $\Pi_O^{sg}$. The results, shown in Figure~\ref{exp:vGraph}.(a), indicate that $\Pi_O^{sg}$ consistently achieves significantly lower update time than the original $\Pi_O$ across all tested values of $L$. Moreover, the update time is nearly independent of graph size and primarily determined by $L$: smaller values of $L$ lead to lower update time costs.

\para{Causality Relation Proof}.  We evaluate the execution time of causality relation proof operations based on the incoming and outgoing path digests ($\Pi_I$ and $\Pi_O$). To assess the time across different graph structures, we construct two graph instances using data from OpTC and StreamSpot. For each graph, we generate forward and backward causality relation proofs for 1,000 randomly selected nodes and record the proof time. The results, shown in Figures~\ref{exp:vGraph}.(b) and \ref{exp:vGraph}.(c), show that the proof time for both types of proofs scales proportionally with the number of connected nodes.

Notably, forward causality proofs can also be derived from segmented outgoing path digests ($\Pi_O^{sg}$). We therefore additionally measure the corresponding proof time. Recall that each node’s $\Pi_O^{sg}$ captures only components within a segmented tree and does not extend to subsequent trees. Thus, generating a complete proof may require retrieving node proofs of segmented tree roots from the graph accumulator~(\S\,\ref{sec:forward_validation}), potentially making the process slower than using $\Pi_O$ directly. Typically, the proof time is affected by the predefined tree segmentation depth $L$. To evaluate this effect, we vary $L$ and record the corresponding proof time, as shown in Figure~\ref{exp:vGraph}.(c). Interestingly, the proof time using $\Pi_O^{sg}$ is comparable to that of $\Pi_O$, and remains nearly independent of the segmentation depth. This is likely because the overhead of retrieving node proofs constitutes only a small fraction of the total time.

 \para{Causality Relation Validation}. Based on the generated proofs, we validate forward and backward causality relations and record the corresponding validation time. As shown in Figures~\ref{exp:vGraph}.(b) and \ref{exp:vGraph}.(c), the validation time is proportional to the number of validated nodes. Moreover, it remains similar regardless of whether the proofs are generated from segmented or unsegmented outgoing path digests ($\Pi_O^{sg}$ and $\Pi_O$).

\para{Remark.} Figures~\ref{exp:vGraph}.(a)–(c) show that segmented outgoing path digests substantially reduce update time while preserving similar proof and validation time to unsegmented ones. A segmentation depth of $1$ offers the best trade-off, minimizing update time with negligible overhead in proof and validation.

\subsection{Computational Cost of \sys}
\label{sec:exp_overall_cost}
Based on prior experimental results, we evaluate the computational cost of the entire \sys workflow. We configure the underlying verifiable graphs with the optimal setup---using segmented outgoing path digests and a segmentation depth of 1. To assess \sys in practice, we set up two endpoint auditing environments using data from 600 and 1,000 distinct endpoints in StreamSpot and OpTC. In these environments, we measure key \sys operations: event addition, commitment, causality proof generation, and validation.

\begin{figure}[ht]
  \centering
  \subfloat[Event recording time]{\includegraphics[width=1.65in]{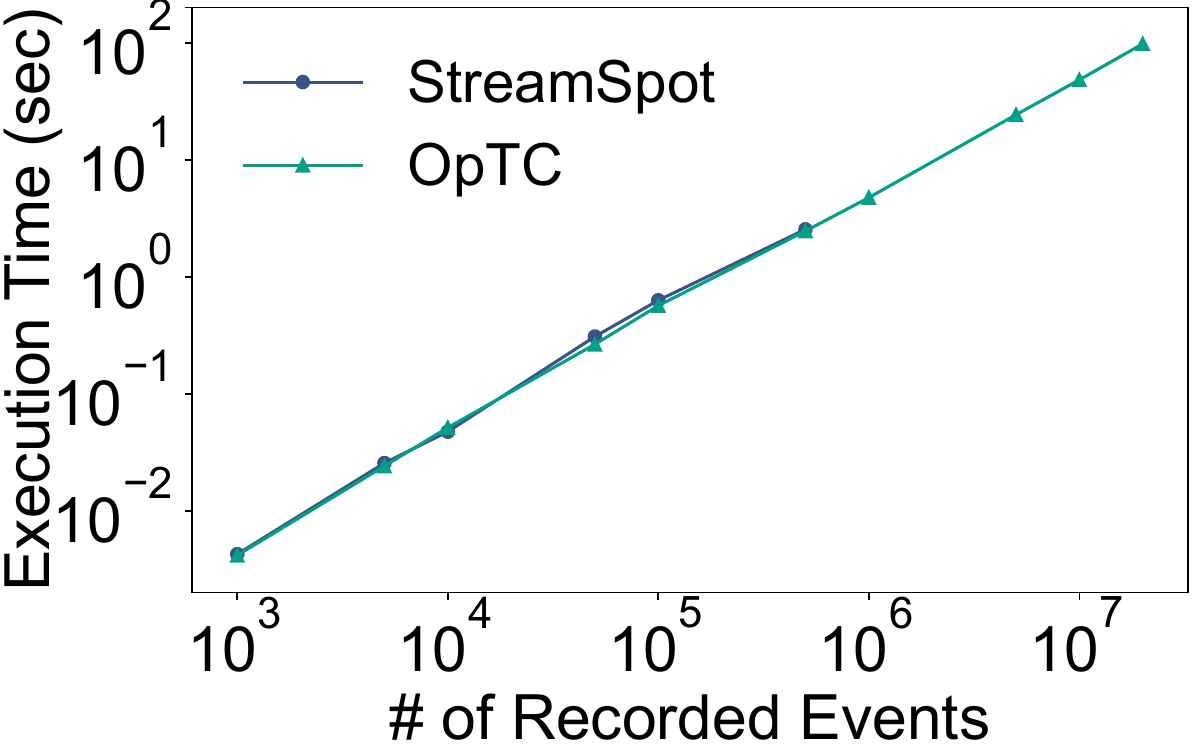}
  }
   \subfloat[Commitment throughput]{\includegraphics[width=1.65in]{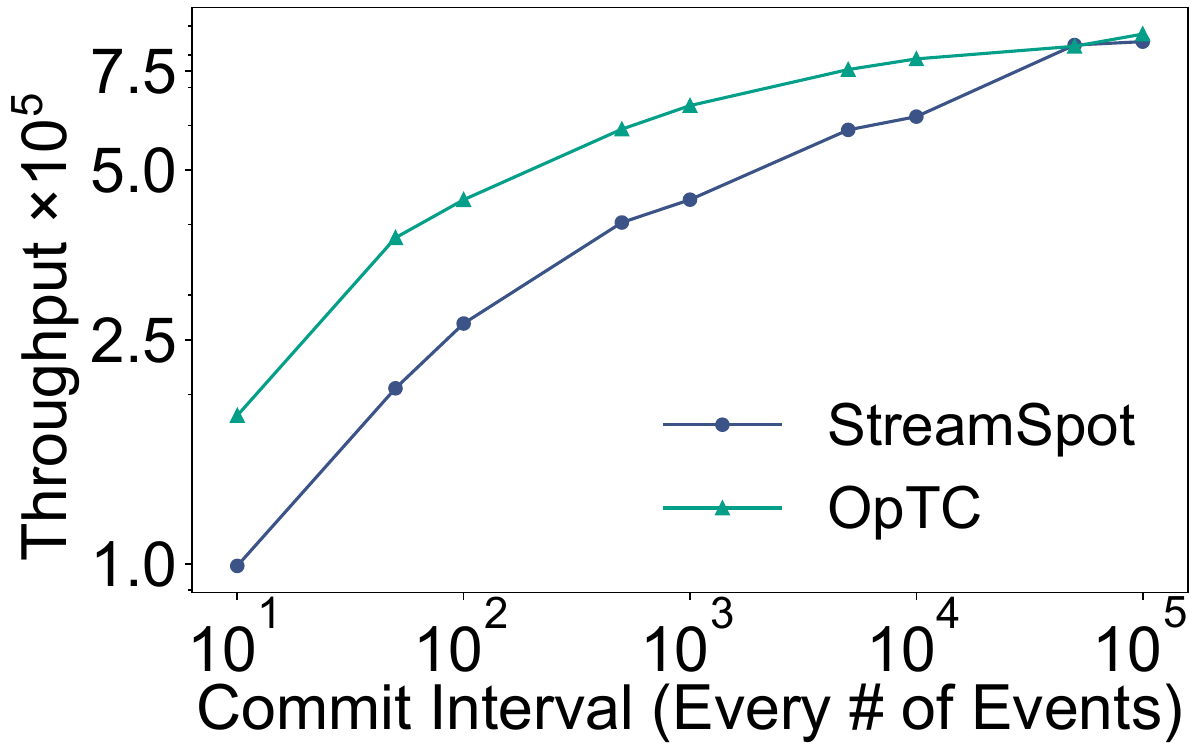}}\\
     \subfloat[Backward causality analysis proof and validation time]{\includegraphics[width=3.3in]{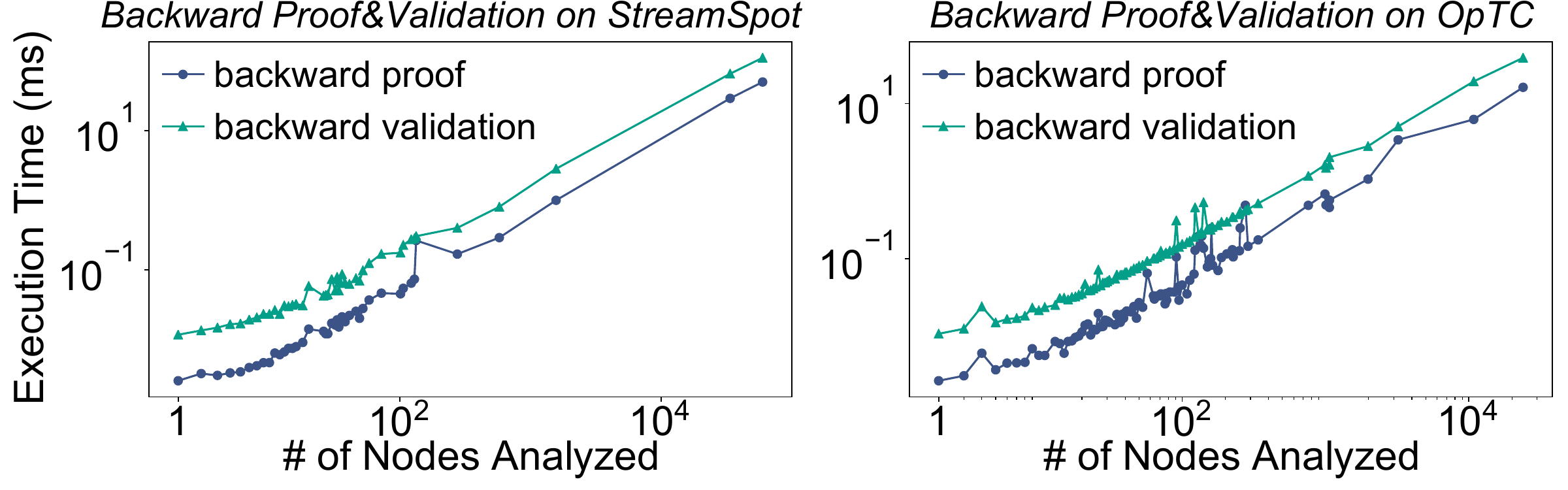}}\\
     \subfloat[Forward causality analysis proof and validation time]{\includegraphics[width=3.3in]{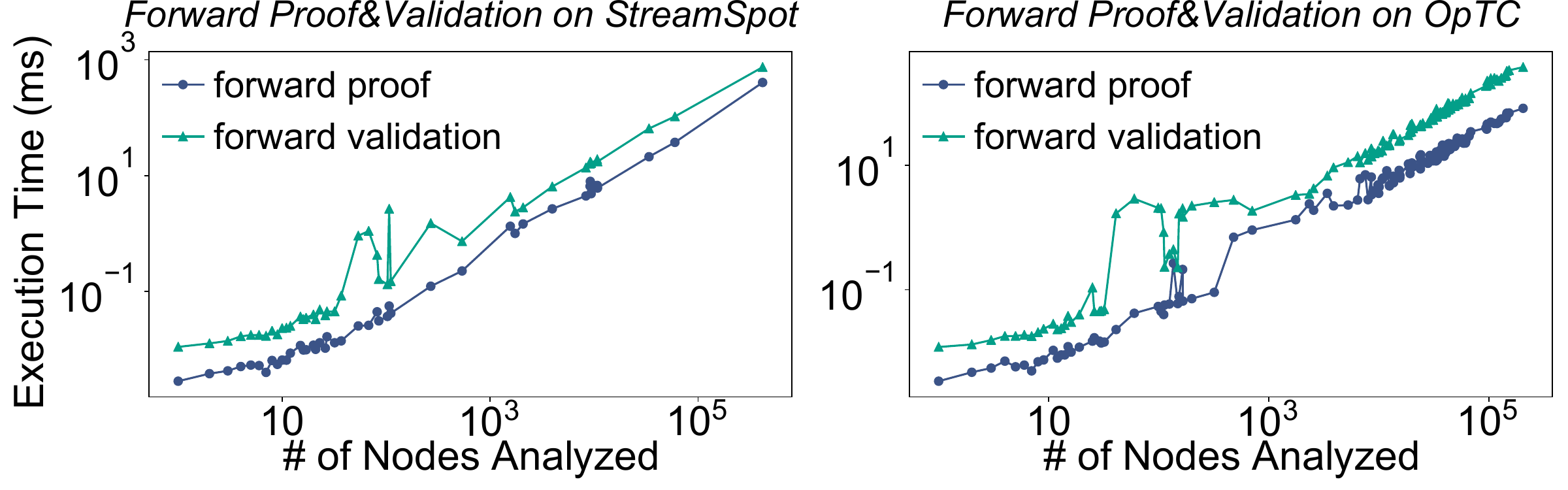}}
   \caption{Computational cost of \sys}
   \label{exp:overall_cost}
\end{figure} 

\para{Event Recording}.  Figure \ref{exp:overall_cost}.(a) shows the average event recording time per endpoint.  The results indicate that the recording time scales linearly with the number of recorded events. Additionally, it remains similar across datasets: for both StreamSpot and OpTC, the per-event recording time ranges from 4 to 6~$\mu$s. Notably, compared to the entire six-day duration of the OpTC dataset, \sys records 25 million events in just 119 seconds---introducing negligible overhead.

\para{Commitment.} 
Recall that each endpoint periodically uses the graph accumulator to commit node changes during event recording (see \S\,\ref{sec:vCause}), enabling timely validation. For endpoints with high event volumes, the commitment frequency affects throughput: overly frequent commitments may redundantly process repeated changes to the same nodes, degrading performance, whereas an appropriate interval allows multiple changes to be merged into a single commitment. To evaluate this, we vary the interval between successive commitments and measure throughput. As shown in Figure~\ref{exp:overall_cost}(b), throughput increases with longer intervals. However, longer intervals may also increase the latency of validating the latest results, indicating a performance trade-off in selecting the interval. Overall, all tested intervals achieve acceptable throughput, ranging from $10^5$ to $8 \times 10^5$ node changes per second.

\para{Causality Analysis Proof and Validation.}
We issue 1,000 random forward and backward causality analysis queries on endpoint logs and record the corresponding proof generation time. As shown in Figures~\ref{exp:overall_cost}.(c) and (d), the time for both forward and backward proofs increases with the number of queried nodes.
Overall, both operations are efficient, \eg generating a forward proof over $10^5$ nodes takes only 49~ms. Based on the generated proofs, we measure the execution time of causality analysis validation. As shown in Figures~\ref{exp:overall_cost}.(c) and \ref{exp:overall_cost}.(d), the validation time also scales proportionally with the number of validated nodes.

\para{Comparison with Enclave-based Validation.}
To demonstrate the efficiency of \sys, we implement an SGX enclave-based causality analysis system and compare it with \sys using 1,000 random forward queries on 2 GB of OpTC endpoint data. To enable the enclave to process and validate logs at this scale, we partition the logs into chunks and generate an HMAC per chunk. The results show the enclave-based approach incurs an average latency of 2408 ms, whereas \sys requires only 3.08 ms. As noted in \S\,\ref{sec:related_validation}, this gap stems from substantial enclave context-switch overhead during full log ingestion, integrity validation, and large-scale graph queries.

\subsection{Runtime Overhead in Practice}
\label{sec:exp_overhead}
To evaluate \sys's runtime overhead in realistic deployments, we integrate it into endpoint systems using two widely used open-source Linux loggers: auditd~\cite{Auditd} and Falco~\cite{neves2018falcon}. At runtime, \sys ingests system event logs from these loggers and processes them in real time. Following prior results, \sys is configured with segmented outgoing path digests (depth 1) and a commitment interval of 1,000 events.

\begin{table}[ht]
\centering
\caption{Runtime overhead of \sys in realistic deployments. We use Apache Bench to issue 10,000 requests with 10 threads on Apache2 and Nginx, and run seven programs from NAS Parallel Benchmarks, reporting the average execution time per request (or run) when auditd is used as the logger.}
\label{tab:overhead}
\small
\begin{tabular}{p{0.7in}p{0.65in}<{\centering}p{0.75in}<{\centering}p{0.55in}<{\centering}}
\toprule
\textbf{Application} & \textbf{Baseline} & \textbf{with \sys} & \textbf{Overhead} \\
\midrule
Apache2 & 17.127 $ms$ & 17.146 $ms$ & 0.11\% \\
Nginx & 15.765 $ms$ & 15.780 $ms$ & 0.10\% \\
NAS-bench & 41.211 $s$ & 41.354 $s$ & 0.35\% \\
\bottomrule
\end{tabular}
\end{table}

Since real-world systems often run applications under high workloads, \sys may introduce performance interference. To evaluate this, we deploy two common web application environments: Apache 2~\cite{Apache} and Nginx~\cite{Nginx}, each hosting a WordPress website~\cite{WordPress}. We then use Apache Bench~\cite{ApacheBench} to issue 10{,}000 requests with 10 threads and measure the average execution time per request. Furthermore, we evaluate compute-intensive workloads from seven programs from the NAS Parallel Benchmarks. As shown in Table~\ref{tab:overhead}, when auditd is used as the logger, \sys introduces less than 1\% overhead compared to the baseline. Similar results are observed with the Falco logger (see Table~\ref{tab:runtime_overhead_falco} in Appendix~\ref{ap:additional_experiments}).

\begin{figure}[t]
  \centering
  \subfloat[Backward causality analysis]{\includegraphics[width=1.65in]{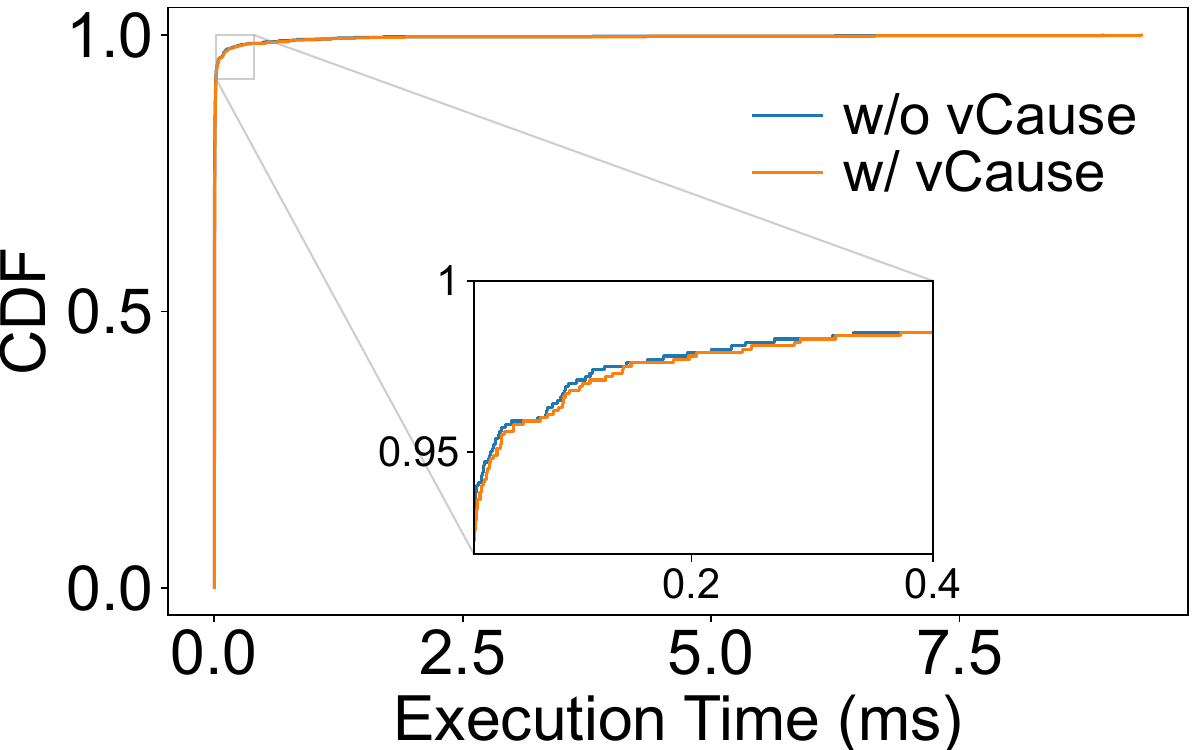}
  }
   \subfloat[Forward causality analysis]{\includegraphics[width=1.65in]{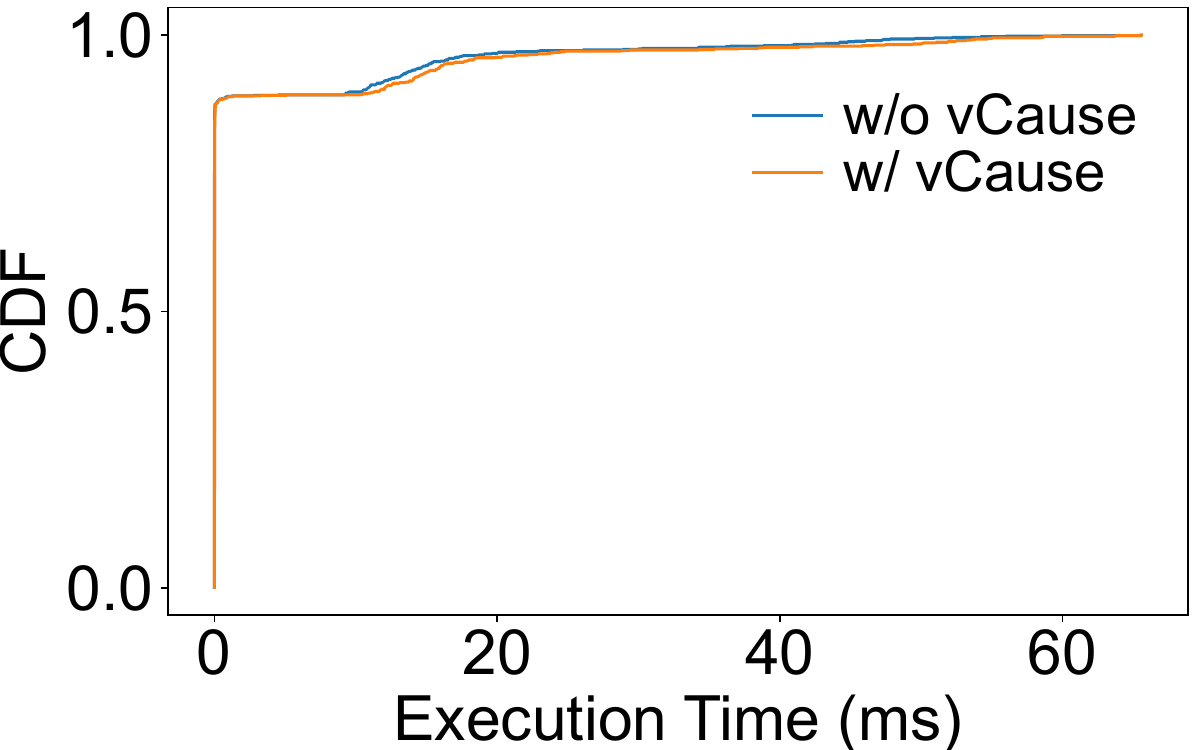}}
   \caption{CDF of causality analysis time}
   \label{exp:cdf_proof_overhead}
\end{figure}

Based on endpoint logs generated by the benchmarking tool, we further evaluate the overhead of cloud-side causality analysis. For comparison, we implement a baseline causality analysis system without validation. Both systems process 1{,}000 randomly selected queries, and we record their execution time. Figure~\ref{exp:cdf_proof_overhead} shows the cumulative distribution function (CDF) of causality analysis time, indicating that \sys incurs only minor overhead while achieving the same proportion of queries. On average, the overhead of forward and backward proof generation is 3.36\% and 8.92\%, respectively.

\subsection{Communication and Storage Costs}
\label{sec:exp_storage_comm}

\begin{figure}[ht]
  \centering
  \subfloat[Communication between each endpoint and cloud]{\includegraphics[width=3.3in]{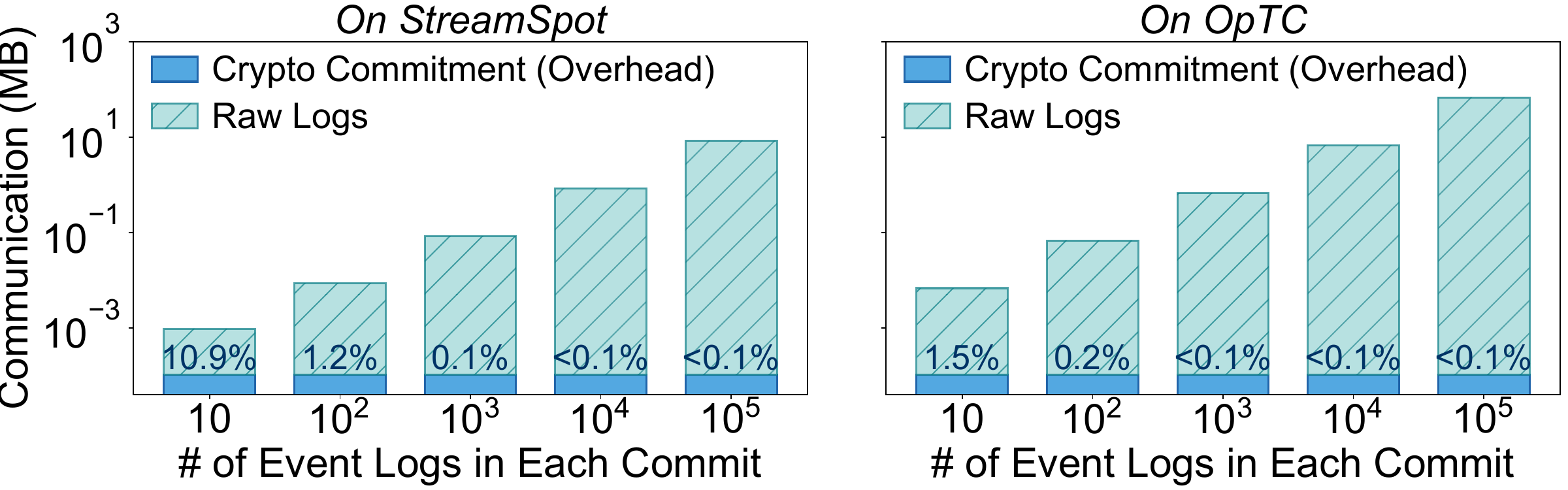}
  }
  \\
  \subfloat[Communication between cloud and administrator]{\includegraphics[width=3.3in]{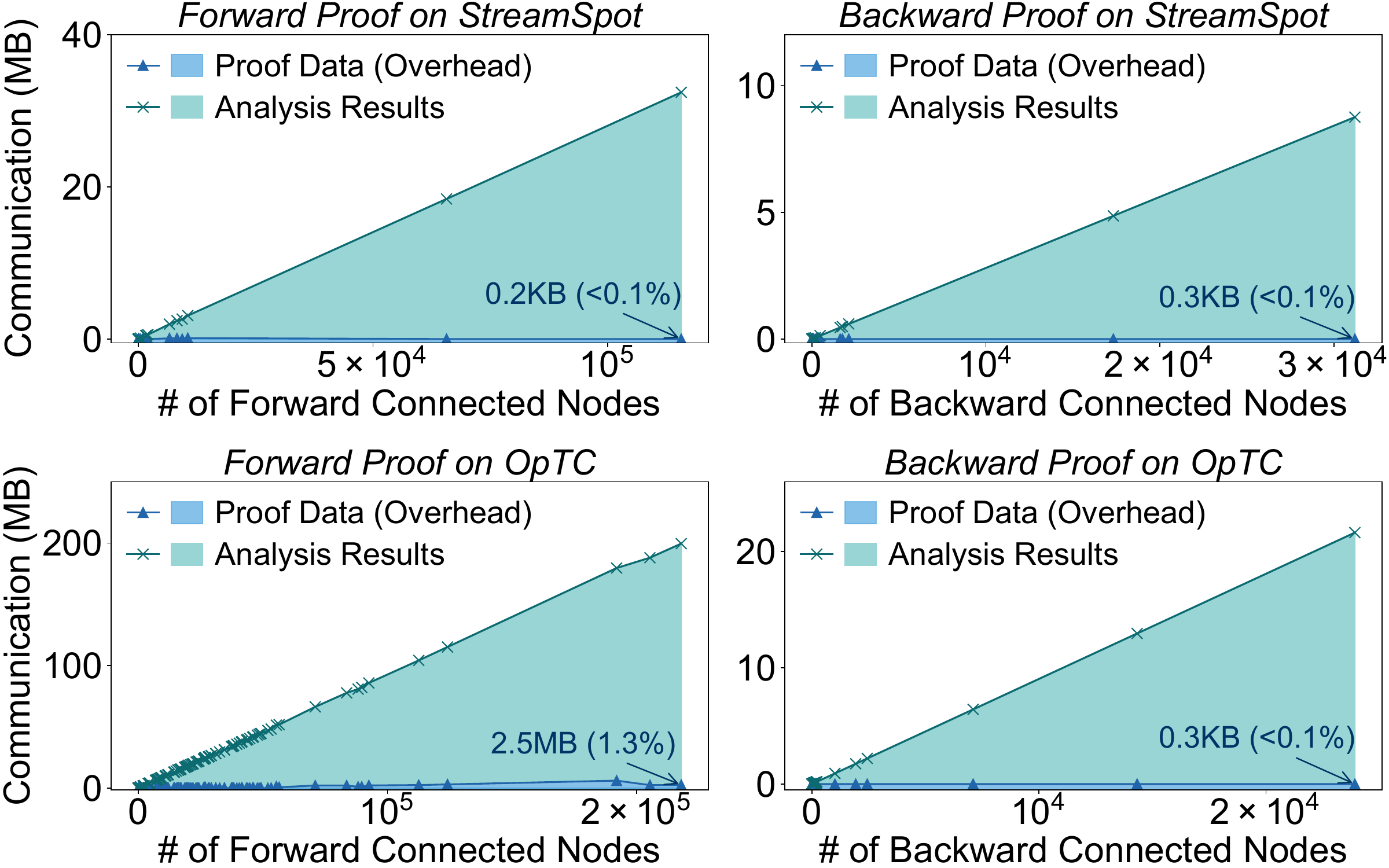}
  }
\caption{Communication cost of \sys}
\label{exp:communication}
\end{figure} 

We run \sys on the StreamSpot and OpTC datasets to evaluate its communication and storage costs at scale. We first measure the communication cost between endpoints and the cloud, focusing on the size of cryptographic commitments (\ie the signed Merkle tree root and associated metadata) transmitted during each commitment. As shown in Figure~\ref{exp:communication}.(a), this cost remains constant at roughly 108 bytes, introducing negligible overhead relative to raw log transmission---below 1.2\% when more than $10^2$ logs are sent.

We also evaluate the communication cost between the cloud and the administrator, which includes causality analysis results and the additional proof data introduced by \sys. Specifically, we simulate 1,000 random forward and backward queries and record the corresponding costs. As shown in Figure~\ref{exp:communication}.(b), the proof data incurs only minor overhead relative to the returned results. Particularly, the proof size grows slightly with the number of returned nodes. This is because only the proof for partial nodes and associated path digests need to be returned for validation (see \S\,\ref{sec:vCause}).

\begin{table}[ht]
\caption{Average storage cost per endpoint. ADS refers to the authenticated data structures introduced by \sys. }
\small
\label{tb:storage_cost}
\begin{tabular}{p{0.6in}p{1.05in}<{\centering}p{0.45in}<{\centering}p{0.55in}<{\centering}}
\toprule
\bf{Dataset}    & \bf{Original Graph} & \bf{ADS}     & \bf{Overhead} \\ \midrule
StreamSpot & 41.85 MB         & 8.29 MB & 19.81\%  \\ \midrule
OpTC       & 15.76 GB         & 1.10 GB & 6.98\%   \\ \bottomrule
\end{tabular}
\end{table}

To assess \sys's storage cost, we measure the size of the authenticated data structures (\ie graph accumulators and path digests of nodes) stored in the cloud. Since forward and backward validations are always performed jointly on provenance graphs, we apply an optimization that merges each node's incoming and outgoing digests into a single digest. 
Table~\ref{tb:storage_cost} reports the average storage per endpoint. Compared to the original graph data, \sys introduces only modest overhead from authentication data. For the OpTC dataset, the overhead is 6.98\%, which we view as more representative of practical storage requirements, since OpTC captures real-world, system-wide endpoint logs, while StreamSpot primarily contains logs collected in a controlled lab environment.

\section{Discussion}
\label{sec:discussion}
\para{Integration with Causality Analysis Variants.} \sys focuses on providing validation for the general form of causality analysis, while also supporting integration with advanced variants that prune or prioritize analysis paths~\cite{ma2016protracer,liu2018towards,hassan2019nodoze,fang2022back,li2023nodlink}. Notably, \sys's causality analysis results encompass the outputs of these variants. Thus, users can use \sys to first verify the complete subgraph related to a point-of-interest node and then pass it to downstream methods for refinement. To further reduce overhead, endpoints can selectively include only critical components in the verifiable graph.
\section{Conclusion}
This paper introduces \sys, an efficient verifiable causality analysis system for cloud-based endpoint auditing. It employs two tailored authenticated data structures: a verifiable versioned provenance graph and a graph accumulator. The graph accumulator enables proof generation for point-of-interest node queries, while the provenance graph uses incoming and outgoing path digests to provide proofs for associated causality relations. Security analysis and experimental results demonstrate the security and efficiency of \sys.
\appendix
\section*{Acknowledgments}

We are grateful to all the anonymous reviewers for their insightful comments, which have greatly improved our paper. The research is supported in part by the National Natural Science Foundation of China (No. 62202465), the National Key Research and Development Program of China (No. 2021YFB2910109), the Beijing Key Laboratory of Network Security Protection Technology (No. 2022YFB3103900), and the Outstanding Talent Scheme (Category B) - Qihang Zhou (E3YY141116).
\section*{Ethical Considerations}
\para{Stakeholders and Potential Impacts.}
This work proposes a verifiable causality analysis framework for cloud-based endpoint auditing.
In this context, the stakeholders include endpoint users whose activities generate audit logs, enterprise security operators, cloud service providers, and the broader security and academic communities.
The expected impacts of this work are enhanced integrity, transparency, and trustworthiness in cloud-based endpoint auditing, leading to greater accountability for service providers and stronger resilience against malicious behaviors.

\para{Dual-Use Considerations and Mitigations.}
While the framework is designed for defensive auditing, verifiable provenance-based causality tracking could, in principle, be misused for intrusive monitoring if deployed without proper safeguards.
To mitigate such risks, all experiments were conducted in controlled lab environments and used publicly available datasets that contain no sensitive raw user information.
The system design adheres to the principle of least privilege and collects only the necessary system-call logs required for verification.
In deployment, endpoint loggers can be configured with privacy-aware policies to control the scope of log collection.

\para{Ethical Reasoning for Publication.}
With reasonable configurations—such as limited log collection and privacy controls---the framework poses low risk while significantly improving transparency and accountability in cloud-based endpoint auditing. We believe its publication can support community efforts toward more trustworthy and responsible auditing.

\section*{Open Science}
To comply with open science principles, we make our code and associated materials publicly available through open repositories. In particular, our prototype implementation, data preprocessing scripts, and test datasets are accessible at \url{https://doi.org/10.5281/zenodo.17908629}, enabling others to review, build upon our work, and validate the associated results.

\bibliographystyle{plain}
\bibliography{ref}
\appendix
\section{Proofs for Node Queries}

\begin{algorithm}[ht]
\SetKwFunction{GlobalSearch}{GlobalSearch}
\SetKwFunction{LocalSearch}{LocalSearch}
\SetKwFunction{Append}{Append}
\SetKwProg{Fn}{Function}{:}{}
\SetKwComment{Comment}{$\triangleright$\ }{}
\DontPrintSemicolon
  \caption{Search and Proof in Hierarchical Trees}
  \label{alg:treesearch}
  \KwIn{Global and local tree roots $r_G$, $r_L$, the queried entity $s$ and temporal relation $r \in \{'\preceq t', '\succeq t'\}$, and the proof ($\rho_G$, $\rho_L$).}
  \KwOut{A boolean value and the matched node.}
    \Fn{\GlobalSearch{$r_G, s, \rho_G$}}{
        \If{$r_G$ is a leaf node}{
            \If{$r_G$.leaf.min $== s$}{
            \Return True, $r_G$; \Comment*[r]{exact matching}
            }
            \Return False, $r_G$;
        }
        \If{$s \in [r_G$.left.min, $r_G$.left.max$]$}{
        $\rho_G$.\Append{$r_G$.right}; \\
        \Return \GlobalSearch{$r_G$.left, $s, \rho_G$};
     } 
     \ElseIf{$s \in [r_G$.right.min, $r_G$.right.max$]$}{
        $\rho_G$.\Append{$r_G$.left};\\
        \Return \GlobalSearch{$r_G$.right, $s, \rho_G$};
     }
     \Else(\hfill\Comment*[h]{not in left and right subtrees}){
        $\rho_G$.\Append{$r_G$}; \\
        \Return False, $r_G$;
     }
    }
  \Fn{\LocalSearch{$r_L, r, \rho_L$}}{
        \If{$r_L$ is a leaf node}{
            \Return True, $r_G$; \Comment*[r]{fuzzy matching}
        }
        \If{($r$ = $\preceq t$ \KwAnd $t \in [r_L$.left.min, $r_L$.right.min$)$) \KwOr ($r$ is $\succeq t$ \KwAnd $t \leq r_L$.left.max)}{
        $\rho_L$.\Append{$r_L$.right};\\
        \Return \GlobalSearch{$r_L$.left, $r, \rho_L$};
     } 
     \ElseIf{$r$ is $\preceq t$ \KwAnd $t \geq r_L$.right.min \KwOr ($r$ is $
    \succeq t$ \KwAnd $t \in (r_L$.left.max, $r_L$.right.max$]$)}{
        $\rho_L$.\Append{$r_L$.right};\\
        \Return \GlobalSearch{$r_L$.left, $r, \rho_L$};
     }
     \Else(\hfill\Comment*[h]{all subtrees not satisfy $r$}){
     $\rho_L$.\Append{$r_L$}; \Return False, $r_L$;
     }
    }
\end{algorithm}

\subsection{Single-Node Proof}
\label{ap:single_node_proof}
\para{Membership Proof.} 
For a single-node query $N(s, r)$, the membership proof is generated by searching the global tree for the system entity keyword $s$ and the local tree for the temporal relation $r$. These searches are implemented by two recursive functions, \emph{GlobalSearch} and \emph{LocalSearch}, as shown in Algorithm~\ref{alg:treesearch}. \emph{GlobalSearch} performs an exact search using $s$, while \emph{LocalSearch} conducts a fuzzy search based on the temporal relation $r \in \{\preceq t, \succeq t\}$.

\para{Non-Membership Proofs.} A non-membership proof for a node query $N(s, r)$ is represented as a tuple $(\tilde{\rho}_G, \tilde{\rho}_L)$, where $\tilde{\rho}_G$ is obtained by searching the global tree for the entity keyword $s$, and $\tilde{\rho}_L$ is obtained by searching the corresponding local tree for the temporal relation $r$. Two cases may arise:

\phreset

\phitem{Condition \#1.}
No node matches $s$, and the search terminates in the global tree. The proof contains only $\tilde{\rho}_G$, consisting of the sibling nodes along the search path.

\phitem{Condition \#2.}
Nodes matching $s$ exist, but none satisfy $r$. The search reaches the local tree of $s$ but terminates without a match for $r$. The proof contains both $\tilde{\rho}_G$ and $\tilde{\rho}_L$, each comprising the sibling nodes encountered in the global and local trees, respectively.

\subsection{Entity-Temporal Range Proof}
\label{ap:temporal_range_proof}
We now extend our graph accumulator to provide proofs for temporal range queries associated with an entity, \ie $Q(s, [a, b])$, where $s$ denotes the entity ID and $[a, b]$ indicates a closed timestamp range. 

\para{Entity-Temporal Range Proofs.}  The proofs are structured as a tuple $(\rho_G, (\rho_{L}^l, \rho_{L}^r))$, where $\rho_G$ is a proof derived from the global tree, and $(\rho_{L}^l, \rho_{L}^r)$ represent range proofs from the corresponding local tree. Specifically, $\rho^l_L$ and $\rho^r_L$ together prove the integrity of successive versioned nodes spanning the temporal range $[a, b]$. Notably, when $a = b$, the proof $(\rho_G, (\rho_{L}^l, \rho_{L}^r))$ becomes equivalent to a single-node proof.

\begin{figure}[ht]
    \centering
    \includegraphics[width=3.3in]{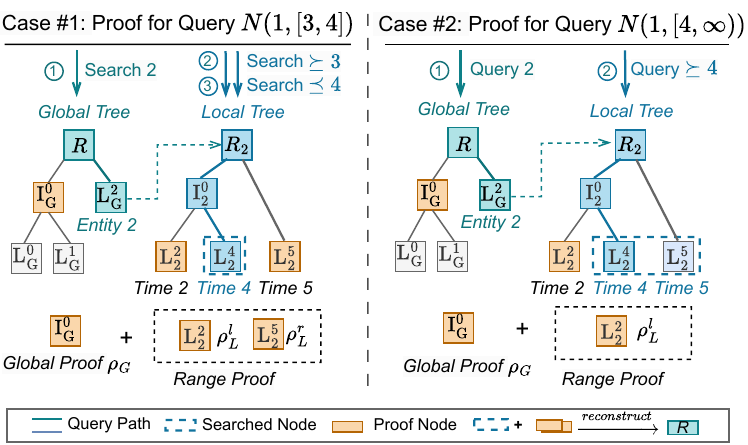}
    \caption{Temporal range proof}
    \label{fig:temporal_range_proof}
\end{figure}

Figure~\ref{fig:temporal_range_proof} illustrates the generation of the proof tuple $(\rho_G, (\rho_{L}^l, \rho_{L}^r))$ for query $Q(s, [a, b])$. The proof $\rho_G$ is generated through an exact search in the global tree. Siblings encountered along the search path are included in $\rho_G$. The local range proofs $\rho_{L}^l$ and $\rho_{L}^r$ are derived from separate searches in the local tree of $s$ for the leftmost and rightmost nodes whose timestamp keywords are within the range $[a, b]$. The left-hand siblings along the search path to the leftmost are integrated into $\rho_{L}^l$, while the right-hand siblings along the path to the rightmost are incorporated into $\rho_{L}^r$.

\begin{algorithm}[t]
\SetKwFunction{GlobalSearch}{GlobalSearch}
\SetKwFunction{LocalSearch}{LocalSearch}
\SetKwFunction{Append}{Append}
\SetKwFunction{NextLeaf}{NextLeaf}
\SetKwProg{Fn}{Function}{:}{}
\SetKwComment{Comment}{$\triangleright$\ }{}
\DontPrintSemicolon
  \caption{{Proving Nodes for Entity-Temporal Range Query $\mathbf{Q(s, [a, b])}$}}
  \label{alg:temporal_range_proof}
  \KwIn{The root  $r$ of the global tree, the queried system entity $s$, the root $r_s$ of system entity $s$'s local tree, and the queried timestamp range $[a, b]$.}
  \KwOut{A boolean value indicating node existence, and the temporal range proof $(\rho_G, (\rho_L^l, \rho_L^r))$.}
  $\rho_G = []$;\\
  $st_s, n_s$ = \GlobalSearch{$r, s, \rho_G$};\\
  \If{$st_s == $ False}{
  \Return \emph{False}, $(\rho_G, (\varnothing, \varnothing))$;
  }
$\rho_L^l = [], \rho_L^r = [];$\\
$st_a$, $lmost$ = \LocalSearch{$r_s, '\succeq a', \rho_L^l$};\\
$st_b$, $rmost$ = \LocalSearch{$r_s, '\preceq b', \rho_L^r$};\\
$\rho_L^l = LeftSib(\rho_L^l)$, $\rho_L^r = RightSib(\rho_L^r)$;\\
\If{$rmost.time > a$ \KwOr $lmost.time < b$}{
    \Return False,$(\rho_G, (\rho_L^l, \rho_L^r))$; 
}
\Return True, $(\rho_G, (\rho_L^l, \rho_L^r))$;
\end{algorithm}

For a query $Q(s,[a,b])$, the corresponding proof generation algorithm is shown in Algorithm~\ref{alg:temporal_range_proof}, which leverages the \emph{GlobalSearch} and \emph{LocalSearch} described in Algorithm~\ref{alg:treesearch}.

\para{Validation.} Given a temporal range proof $(\rho_G,(\rho_L^l,\rho_L^r))$ and the queried nodes, the verifier reconstructs the search path and tree structure. The proof is valid if the reconstructed root matches the committed root. To verify that the queried nodes satisfy $Q(s,[a,b])$, the verifier checks that they lie in the local tree of $s$ and that their timestamp keywords fall within $[a,b]$. Notably, using temporal range proofs enables efficient batch verification of $m$ nodes. Verifying nodes individually incurs $O(m\log N)$ complexity, whereas batch verification with temporal range proofs reduces this cost to $O(m)$.

\begin{algorithm}[ht]
\SetKwFunction{TreeMerge}{TreeMerge}
\SetKwFunction{Append}{Append}
\SetKwFunction{Empty}{Empty}
\SetKwFunction{Pop}{Pop}
\SetKwFunction{Push}{Push}
\SetKwProg{Fn}{Function}{:}{}
\SetKwComment{Comment}{$\triangleright$\ }{}
\DontPrintSemicolon
  \caption{{\dtree Node Insertion Operation}}
  \label{alg:dimtree_node_insertion}
  \KwIn{A new leaf node $t$, the leaf node vector $L$, and a stack storing previous subtree roots $s$. }
    \KwOut{a new stack $s$. }

        $L$.\Append{$t$};\\
        \While{$s$.\Empty{}}{
            \If{$s.top.height \neq t.height$}{
                \Break;
            }
            $t$ = \TreeMerge{$s.top, t$};\\
            $t.height += 1$; $s$.\Pop{};
        }
    $s$.\Push{$t$};\\
    \Return $s$;
\end{algorithm}

\section{Node Insertion Algorithm in \dtree}
\label{ap:node_insertion_algorithm}
Algorithm \ref{alg:dimtree_node_insertion} shows the node insertion in \dtree.

\section{Algorithms of Causality Relation Validation}

\begin{algorithm}[ht]
\SetKwFunction{RDFS}{RDFS}
\SetKwFunction{Add}{Add}
\SetKwFunction{Hash}{H}
\SetKwFunction{DFSInDigest}{DFSInDigest}
\SetKwFunction{Main}{Main}
\SetKwProg{Fn}{Function}{:}{}
\SetKwComment{Comment}{$\triangleright$\ }{}
\DontPrintSemicolon
  \caption{Validating Backward Causality Relations for a Given Node}
  \label{alg:backward_verification}
  \KwIn{a node $n$, its backward causally related nodes and edges $\{V_{\rightarrow n}, E_{\rightarrow n}\}$, and the hash function \Hash{$\cdot$}.}
  \KwOut{a boolean variable.}
\Fn{\Main{$n$, $\{V_{\rightarrow n}, E_{\rightarrow n}\}$}}{
$original = n.\Pi_I$;\\
\DFSInDigest{$n$, $\{V_{\rightarrow n}, E_{\rightarrow n}\}$, $\{\}$};\\
\If{$n.\Pi_I == original$}{
    \Return \emph{True};
}
\Return \emph{False};\\
}
 \Fn{\DFSInDigest{$n, \{V_{\rightarrow n}, E_{\rightarrow n}\}, visited$}}{
    \If{$n \in visisted$}{
        \Return;
    }
    $visited$.\Add{n};\\
    $byte = ''$;\\
    \For{$e \in n.E_{in}$}{
        \If{$e \notin E_{\rightarrow n}$ \KwOr $e.src \notin V_{\rightarrow n}$}{
            \Return;
        }
        \DFSInDigest($e.src$);\\
        $byte~|| =  str(e)~||~e.src.\Pi_I$;\\
    }
    $n.\Pi_{I}$ = \Hash{$byte$};
}
\end{algorithm}
 Algorithm~\ref{alg:backward_verification} and \ref{alg:forward_verification} detail the process of validating a node $n$'s forward and backward causality relations using incoming and segmented outgoing path digests.

\begin{algorithm}[ht]
\SetKwFunction{DFS}{DFS}
\SetKwFunction{Add}{Add}
\SetKwFunction{Hash}{H}
\SetKwFunction{ExtractRoot}{ExtractRoot}
\SetKwFunction{NodeVerify}{NodeVerify}
\SetKwFunction{DFSOutDigest}{DFSOutDigest}
\SetKwProg{Fn}{Function}{:}{}
\SetKwComment{Comment}{$\triangleright$\ }{}
\DontPrintSemicolon
  \caption{Validating Forward Causality Relations for a Given Node}
  \label{alg:forward_verification}
  \KwIn{a node $n$, its forward causally related nodes and edges $\{V_{ n \rightarrow}, E_{ n \rightarrow}\}$, temporal range node proofs $P$, and the hash function \Hash{$\cdot$}.}
  \KwOut{a boolean variable.}
    \Fn{\Main{$n$, $\{V_{n\rightarrow}, E_{n\rightarrow}\}$}}{
        $rts$ = \ExtractRoot{$V_{ n \rightarrow}$};\\
        \NodeVerify{$rts$, $P$};\\
        \ForEach{$s \in \{n\} \bigcup rts  $}{
            $original = s.\Pi_O^{sg}$;\\
            \DFSOutDigest{$s$, $\{V_{\rightarrow n}, E_{\rightarrow n}\}$, $\{\}$};\\
            \If{$s.\Pi_O^{sg} \neq original$}{
                \Return \emph{False};
            }       
        }
        \Return \emph{True};\\ 
    }
            
 \Fn{\DFSOutDigest{$n,\{V_{n\rightarrow}, E_{n\rightarrow}\},visit$}}{
    \If{$n \in visit$}{
        \Return;
    }
    \If{$n$ is a terminal node}{
        \Return;
    }
    $visit$.\Add{n};\\
    $byte = ''$;\\
    \For{$e \in n.E_{out}$}{
        \If{$e \notin E_{n\rightarrow}$ \KwOr $e.dst \notin V_{n\rightarrow}$}{
            \Return;
        }
        \DFSOutDigest($e.dst$);\\
        $byte~|| =  str(e)~||~e.dst.\Pi_O^{sg}$;\\
    }
    $n.\Pi_{O}^{sg}$ = \Hash{$byte$};
}
\end{algorithm}
\section{Detailed Security Definitions and Proofs}
\label{ap:sec_analysis}

\subsection{Security of Verifiable Versioned Provenance Graphs}
\label{ap:graph_security}
We follow the preimage resistance property of hash functions to define the key security property of our verifiable versioned provenance graphs, \ie causality relation unforgeability. It is formalized via the following game $Forge_{\mathcal{A}}^{vGraph}(k)$:
\phreset
    \phitem{Setup.} The challenger initializes system parameters.
    \phitem{Challenge.} The challenger selects a node $n$ and sends its backward and forward causally related nodes and edges $\{V_{\rightarrow n}, E_{\rightarrow n}\}$ and $\{V_{n \rightarrow}, E_{n \rightarrow}\}$ to $\mathcal{A}$.
    \phitem{Forgery.} The adversary $\mathcal{A}$ attempts to add, delete, or modify a node and edge in either $\{V_{\rightarrow n}, E_{\rightarrow n}\}$ or $\{V_{n \rightarrow}, E_{n \rightarrow}\}$, then sends them back to the challenger.
    
\phitem{Validation.} The challenger examines $n$'s forged  causally related nodes and edges, $\{V_{\rightarrow n}^*, E_{\rightarrow n}^*\}$ and$\{V_{n \rightarrow}^*, E_{n \rightarrow}^*\}$, using $n$'s incoming and outgoing path digests, respectively. If the validation succeeds, $\mathcal{A}$ wins the game and outputs 1.

\begin{myDef}[Causality Relation Unforgeability]
A verifiable versioned provenance graph achieves causality relation unforgeability if, for any polynomial-time adversary $\mathcal{A}$:
\begin{equation}
    \Pr(Forge_{\mathcal{A}}^{vGraph}(k) = 1) \leq \text{negl}(k)
\end{equation}
where $\text{negl}(k)$ denotes a negligible function.
\end{myDef}

Now, we give the following security theorem and proofs.
\begin{myTheo}
A verifiable versioned provenance graph achieves causality relation unforgeability if the used hash function is second-preimage resistant. 
\end{myTheo}

\para{Proof.}
\emph{Case \#1:} Suppose the adversary $\mathcal{A}$ adds, deletes, or modifies a node or edge within the backward causality set ${V_{\rightarrow n}, E_{\rightarrow n}}$. If there exists a path of length $l$ from the queried node $n$ to the tampered node or edge, then there should exist at least one intermediate node $n_i$ along the path for which the following condition holds:

Let the original incoming path string of $n_i$ be defined as:
\[ m = \{(s(e) \parallel e.src.\Pi_I) \mid e \in n_i.E_{in}\} \]
After tampering, the string becomes $m^*$. To win the game, $\mathbf{A}$ should ensure that the resulting hash digest $H(m^*)$ equals the original digest $n_i.\Pi_I = H(m)$, in order to pass the validation described in \S\,\ref{sec:backward_validation}. Given this equality, the adversary $\mathcal{B}$ can construct a second-preimage attack by outputting the pair $(m, m^*)$ such that $m \neq m^*$ and $H(m) = H(m^*)$.

\emph{Case \#2:} If the added, deleted, or modified node or edge is within $\{V_{n \rightarrow}, E_{n \rightarrow}\}$, the construction for $\mathcal{B}$ is similar to the first case. That is, the modified outgoing path digest $n_i.\Pi_O$ after tampering yields another second-preimage attack pair $(m, m^*)$ satisfying $H(m) = H(m^*)$ and $m \neq m^*$.

In conclusion, if the hash function is second-preimage resistant---meaning that the probability of such an attack succeeding is negligible---then the verifiable versioned provenance graph achieves causality relation unforgeability. \hfill \(\square\)

\subsection{Security of \sys}
We now present the security proof for \sys's core security property, \ie the unforgeability of causality analysis results (Theorem~\ref{theo:vCause_sec}). For simplicity, we focus on \sys with unsegmented outgoing path digests. The analysis naturally extends to the segmented setting, where validation involves multiple analogous sub-validations over segmented trees.

\para{Proof for Theorem~\ref{theo:vCause_sec}.}
We prove the unforgeability of causality analysis results via a standard hybrid argument over four games, $G_0$–$G_3$, where $G_0$ refers to the original $Forge_{\mathcal{A}}^{\sys}$ game. Each subsequent game modifies the previous one using a simulator $S$ that emulates the signature scheme, Merkle tree, and verifiable versioned provenance graph.

\phreset
\phitem{Game $G_0$.}
$G_0$ is equivalent to the original $Forge_{\mathcal{A}}^{\sys}$ game, except that it immediately outputs 1 (a win for the adversary) whenever a successful forgery is detected, regardless of which cryptographic component is compromised. Hence,
\begin{equation}
Pr[Forge_{\mathcal{A}}^{\sys}(k) = 1] \leq Pr[G_0 = 1].
\end{equation}
From $\mathcal{A}$, we can construct three adversaries $\mathcal{B}_1$, $\mathcal{B}_2$, and $\mathcal{B}_3$ that attack the signature, Merkle tree, and provenance graph.

\phitem{Game $G_1$.}
This game is identical to $G_0$, except that $S$ now simulates the signature scheme. Verification on $(R,t,\sigma_R)$ is replaced by checking bookkeeping records; unmatched or invalid signatures are rejected. As $S$ perfectly simulates signing, $\mathcal{B}_1$ has negligible advantage. Hence, comparing $G_0$ and $G_1$, we have:
\begin{equation}
|Pr[G_1 = 1] - Pr[G_0 = 1]| \leq Adv*{\mathcal{B}_1}^{sign}.
\end{equation}

\phitem{Game $G_2$.}
$G_2$ is identical to $G_1$, except that $S$ now simulates the Merkle tree. For each node $n$, if proof $\rho_n$ mismatches the recorded data or the queried value, verification rejects. As $S$ can perfectly simulate the Merkle tree behavior, $\mathcal{B}_2$ cannot win a forgery game against the Merkle tree structure. Thus,
\begin{equation}
|Pr[G_2 = 1] - Pr[G_1 = 1]| \leq Adv*{\mathcal{B}_2}^{Merkle}.
\end{equation}

\phitem{Game $G_3$.}
$S$ further simulates the verifiable provenance graph. The validation of node $n$’s causally related components is replaced by a check against bookkeeping records; For node $n$, if its backward or forward components $\{V_{\rightarrow n},E_{\rightarrow n}\}$ or $\{V_{n\rightarrow},E_{n\rightarrow}\}$ mismatch their paths, verification rejects. Hence,
\begin{equation}
|Pr[G_3 = 1] - Pr[G_2 = 1]| \leq Adv*{\mathcal{B}_3}^{vGraph}.
\end{equation}

\phitem{Conclusion.}
By summing up the above results, we have:
\begin{equation} \begin{split} Pr[Forge_{\mathcal{A}}^{\sys} = 1] \leq Pr[G_0 = 1] \leq Adv_{\mathcal{B}_1}^{sign} \\ + Adv_{\mathcal{B}_2}^{Merkle} + Adv_{\mathcal{B}_3}^{vGraph} + Pr[G_3 = 1] \end{split} \end{equation}
All cryptographic components in $G_3$ are perfectly simulated, preventing any forgery; hence $Pr[G_3 = 1] = 0$. The adversary’s overall advantage is therefore negligible, \ie $negl(k)$, assuming the signature scheme is EUF-CMA secure, the Merkle tree is position-binding, and the verifiable versioned provenance graph ensures causality-relation unforgeability.
\hfill $\square$

\section{Additional Evaluation Results}
\label{ap:additional_experiments}

\para{Runtime Overhead of \sys.}
Table~\ref{tab:runtime_overhead_falco} reports the average per-request (or per-run) runtime overhead for Apache2, Nginx, and NAS Benchmarks when Falco is used as the logger.

\begin{table}[ht]
\centering
\caption{Runtime overhead of \sys under realistic deployment settings with Falco as the logger.}
\label{tab:runtime_overhead_falco}
\small
\begin{tabular}{p{0.7in}p{0.65in}<{\centering}p{0.75in}<{\centering}p{0.55in}<{\centering}}
\toprule
\textbf{Application} & \textbf{Baseline} & \textbf{with \sys} & \textbf{Overhead} \\
\midrule
Apache2   & 16.998 $ms$ & 17.018 $ms$ & 0.12\% \\
Nginx     & 15.652 $ms$ & 15.668 $ms$ & 0.10\% \\
NAS-Bench & 41.312 $s$  & 41.489 $s$  & 0.43\% \\
\bottomrule
\end{tabular}
\end{table}

\para{Case Study.}
We take a medium-scale backward/forward causality analysis on OpTC endpoint~660 as an example. The backward analysis involves 19{,}088 nodes and completes in 11.8\,ms, while the forward analysis involves 639 nodes and completes in 0.8\,ms. To evaluate \sys’s verifiability, we modify each node's attributes and run validation. We observe that all tampering attempts are correctly detected. On average, validation takes 1.7\,ms for forward analysis and 31.0\,ms for backward analysis.

\end{document}